\pdfoutput=1 

\documentclass[a4paper, 11pt, final]{article}

\usepackage{amssymb} 
\usepackage{amsthm} 
\usepackage{amsmath,empheq}
\usepackage{bbm}

\DeclareMathAlphabet{\mathcal}{OMS}{cmsy}{m}{n}
\DeclareMathAlphabet\mathbfcal{OMS}{cmsy}{b}{n}

\DeclareFontFamily{U}{dutchcal}{\skewchar\font=45 }
\DeclareFontShape{U}{dutchcal}{m}{n}{<-> s*[1.0] dutchcal-r}{}
\DeclareFontShape{U}{dutchcal}{b}{n}{<-> s*[1.0] dutchcal-b}{}
\DeclareMathAlphabet{\mathcald}{U}{dutchcal}{m}{n}
\SetMathAlphabet{\mathcald}{bold}{U}{dutchcal}{b}{n}
\DeclareMathAlphabet\mathcalz{T1}{pzc}{mb}{it}

\usepackage[nice]{nicefrac} 
\usepackage{siunitx} 

\usepackage{newtxtext, newtxmath} 

\usepackage{anyfontsize}
\usepackage[utf8]{inputenc}
\usepackage[T1]{fontenc}
\usepackage{microtype} 

\providecommand{\JEL}[1]{\textit{\textbf{JEL: }} #1}
\providecommand{\keywords}[1]{\textbf{\textit{Keywords--- }} #1}

\usepackage{setspace} 

\usepackage{parskip} 
\usepackage{etoolbox}
\usepackage{titlesec}
\titleformat{\section}{\normalfont\Large\bfseries}{\thesection}{1em}{}
\titleformat{\subsection}{\normalfont\large\bfseries}{\thesubsection.}{1em}{}
\titleformat{\subsubsection}{\normalfont\normalsize\itshape}{\thesubsubsection.}{1em}{}

\usepackage{abstract}

\renewenvironment{abstract}
 {\normalfont
  \begin{center}
  \bfseries \abstractname\vspace{-.5em}\vspace{0pt}
  \end{center}
  \list{}{
    \setlength{\leftmargin}{0cm}%
    \setlength{\rightmargin}{\leftmargin}%
  }%
  \item\relax}
 {\endlist}

\usepackage{authblk} 
\usepackage[bottom]{footmisc} 

\usepackage{multicol} 

\usepackage{enumitem} 

\usepackage{graphicx} 
\usepackage{float} 
\usepackage{subcaption}
\usepackage{afterpage} 

\usepackage{printlen}

\usepackage[labelfont=bf]{caption}
\usepackage{color, colortbl}
\definecolor{LightGray}{rgb}{0.93,0.914,0.914}    
\usepackage{soul} 
\usepackage{longtable,rotating} 
\usepackage{booktabs} 
\usepackage{multirow} 
\usepackage{arydshln} 
\usepackage{bigdelim} 

\usepackage[most]{tcolorbox}            

\PassOptionsToPackage{hyphens}{url} 

\usepackage{fancyhdr}
\fancyheadoffset{0cm}
\makeatletter
\newcommand{\quickwordcount}[1]{
  \immediate\write18{texcount -quiet -incbib -sub=none -utf8 -1 -sum -merge -encoding=utf8 #1.tex > #1-words}%
  \immediate\openin\somefile=#1-words
  \read\somefile to \@@localdummy
  \immediate\closein\somefile
  \setcounter{wordcounter}{\@@localdummy}
  \@@localdummy
}
\makeatother

\usepackage{silence}
\usepackage[style=apa,backend=biber,natbib,hyperref]{biblatex}
\DeclareLanguageMapping{british}{british-apa}
\defbibenvironment{bibliography}{\enumerate}{\endenumerate}{\item}

\usepackage[colorlinks=true,allcolors=gray]{hyperref} 
\let\orgautoref\autoref

\renewcommand{\autoref}[1]
{%
\def\equationautorefname{Eq.}%
\def\sectionautorefname{Sec.}%
\def\subsectionautorefname{Subsec.}%
\def\figureautorefname{Fig.}%
\def\subfigureautorefname{Fig.}%
\orgautoref{#1}%
}

\usepackage[nameinlink,capitalise]{cleveref}

\usepackage{algorithm,algpseudocode}

\makeatletter
\newlength{\trianglerightwidth}
\algnewcommand{\LineCommentCont}[1]{\Statex \hskip\ALG@thistlm%
  \parbox[t]{\dimexpr\linewidth-\ALG@thistlm}
{\leftskip=\algorithmicindent
  \hangindent=\algorithmicindent 
  \hangafter=1%
  \strut\makebox[\algorithmicindent][c]{$\triangleright$}#1\strut}
  } 
\makeatother

\usepackage[left=1.8cm, right=1.8cm, bottom=2.5cm, top=2.5cm]{geometry}

\usepackage{listings}
\usepackage{xcolor}
\begin{document}


\renewcommand{\figureautorefname}{Fig.}
\onehalfspacing



\newcommand{\MainTitleText}{Comparing two approaches for modelling the loss given default of credit cards: Run-off triangles vs regression}

\title{\fontsize{20pt}{0pt}\selectfont\textbf{\MainTitleText
}}


\author[,a,b]{\large Arno Botha \thanks{ ORC iD: 0000-0002-1708-0153; Corresponding author: \url{arno.spasie.botha@gmail.com}}}
\author[,a]{\large Henko Crewe \thanks{ ORC iD: 0009-0004-3704-6333; email: \url{henko.crewe@gmail.com}}}
\author[,a]{\large Marcel Muller \thanks{ORC iD: 0000-0002-8585-2426; email: \url{marcelcelliers@gmail.com}}}
\author[,a]{\large Janette Larney \thanks{ ORC iD: 0000-0003-0091-9917; email: \url{janette.larney@nwu.ac.za}}}
\affil[a]{\footnotesize \textit{Centre for Business Mathematics and Informatics \& Unit for Data Science and Computing, North-West University, Potchefstroom, South Africa}}
\affil[b]{\footnotesize \textit{National Institute for Theoretical and Computational Sciences (NITheCS), Potchefstroom, South Africa}}
\renewcommand\Authands{, and }

    

\makeatletter
\renewcommand{\@maketitle}{
    \newpage
     \null
     \vskip 1em%
     \begin{center}%
      {\LARGE \@title \par
      	\@author \par}
     \end{center}%
     \par
 } 
 \makeatother
 
 \maketitle

{
    \setlength{\parindent}{0cm}
    \rule{1\columnwidth}{0.4pt}
    \begin{abstract}
    The use of run-off triangles (ROTs) is a common industry practice in estimating the loss given default (LGD) risk parameter when predicting credit losses in banking. We benchmark this industry practice using credit card data against a more sophisticated (though classical) regression-based approach, which is able to leverage various types of input variables in producing loan-level LGD-estimates. This regression-based approach can demonstrably recover the typical characteristics of the `U-shaped' empirical LGD-distribution, which the ROT-based approach cannot do. First, we critically review the ROT-based approach and identify multiple demerits using data-driven diagnostics. We then estimate a two-stage regression-based LGD-model and favourably assess the model performance of each component (or `stage'). Finally, we aggregate the LGD-estimates produced by each approach over time, and compare each time series to the mean empirical loss rate over time. The ROT-based aggregates diverge substantially from the empirical rate over most time periods, whilst the regression-based aggregates follow the empirical trends much closer. These results underscore the greater prediction accuracy of the regression-based LGD-model, relative to the ROT-based one. By implication, the former approach is probably better than the latter ROT-based approach when estimating the LGD under the IFRS 9 accounting framework, which prioritises accuracy.
    \end{abstract}
     
    \keywords{Loss given default (LGD); Write-off; Run-off triangles; Generalised linear models; IFRS 9}
     
     \JEL{C31, C39, G21.}
    
    \rule{1\columnwidth}{0.4pt}
}

\noindent Word count (excluding front matter and appendices):  10,146 

\subsection*{Disclosure of interest and declaration of funding}
\noindent This work is not financially supported by an institution or study grant, and has no conflicts of interest that may have influenced the outcome of this work.



\newpage

\section{Introduction}
\label{sec:intro}

Loss given default (LGD) is an important risk parameter when modelling the credit risk of loans within a portfolio. It generally constitutes the proportion of the default balance that is ultimately lost and unrecoverable. From the perspective of the Basel Capital accords from the \citet{basel2019}, CRE 36.76 therein describes the LGD as an "economic loss". As such, the LGD ought to include all relevant factors, material discount effects over time, and direct/indirect costs that accompany the collection efforts. The Basel capital accords broadly strive towards protecting a bank against unexpected (or catastrophic) losses by specifying rules for reserving a stable level of (regulatory) capital over time. 
In contrast, the \textit{International Financial Reporting Standard} (IFRS) 9 from the \citet{ifrs9_2014} relates to reserving financial provisions to cover the \textit{expected credit losses} (ECLs), or "business-as-usual" losses that result from lending operations. In choosing between these two perspectives (Basel vs IFRS 9), we shall investigate the LGD within the setting of IFRS 9. This setting is chosen particularly since IFRS 9 calls for greater accuracy ("unbiased estimates") than the Basel framework; itself prioritising stability in capital levels over accuracy, as discussed by \citet{crook2010dynamic}.

Whilst not explicitly prescribed by IFRS 9, the LGD is generally structured as a component in a few types of statistical models that inform the ECL of a financial asset (or loan). In turn, the ECL represents the probability-weighted sum of cash shortfalls that are expected to be lost over a specific time horizon, as discussed in paragraphs \S 5.5.17--18, \S B5.5.28--31, and \S B5.5.41--44 of IFRS 9. Having considered a range of possible outcomes that can affect the value of the loan, the estimated ECL-amount should generally be "unbiased", i.e., accurate. This ECL-amount should principally also consider the time value of money, past events, current conditions, and forward-looking information such as macroeconomic forecasts. Furthermore, the ECL-estimation process is typically staged (or staggered) into three stages such that the final amount depends on the degree of deterioration in the credit risk of the asset. As discussed and explored by \citet{botha2025sicr}, and as stated in paragraphs \S 5.5.3--5.5.5 of IFRS 9, each stage necessitates a greater ECL-amount in reflecting an overall pattern of decay (or improvement) in credit quality over time.
Finally, a change in the estimated ECL-amount of a loan between two successive time points is then accounted for in a central loss provision, which should itself ultimately absorb future write-off amounts as they unfold.

The LGD fundamentally represents the \textit{workout} process of a defaulted loan, during which time the collections personnel attempt to nurse the strained relationship between bank and borrower back to health. Several remedial actions exist in furthering this pursuit, which include renegotiating (or reviewing) the contractual loan terms such that the instalment ultimately becomes affordable again; see \citet[pp.~26-28]{VanGestel2009book}, \citet[pp.~11-13]{finlay2010book}, and \citet[\S 2.2]{botha2021Proc}. If these attempts are successful, then the loan is said to have `cured' from default in making a full recovery, and a zero credit loss is commonly assigned. However, if unsuccessful, then the bank typically initiates legal proceedings towards foreclosing on any available collateral, such as property. In so doing, the bank effectively abandons the relationship and the focus shifts entirely to recovering the maximum amount in the shortest period of time. The ensuing receipts (or recoveries) are then offset against the last-known balance of the loan, whereafter any non-zero remainder is written-off as a credit loss. \citet{larney2025} therefore stated that the workout process can only culminate in two possible outcomes: a cured or a written-off outcome. Any remaining cases that are still stuck in the workout process are considered to be unresolved (or right-censored) cases.

There are four common methods by which the LGD-quantity can be estimated, as discussed by \citet{schuermann2004we}, \citet[pp.~217-222]{VanGestel2009book}, and \citet[\S 10]{baesens2016credit}. We shall however focus on the \textit{workout} method that uses a bank's own internal data in estimating the LGD. This method leverages the sequence of collected receipts over the history of a written-off loan, and computes the discounted sum thereof over the workout period back to the default point. The percentage change between this discounted sum and the default balance then constitutes the empirical loss rate, or equivalently, the "actual LGD". The statistical distribution of these loss rates typically have two defining characteristics, as discussed and illustrated by \citet{schuermann2004we}, \citet{calabrese2010bank}, \citet{loterman2012benchmarking}, \citet[\S 10]{baesens2016credit}, \citet{larney2025}, and \citet{botha2026LGD_Surv}. These characteristics include a heavy left-skewed tail, and bimodality in the distribution with one of the modes typically centred at zero as a result of zero-loss cures. In our data, we encounter these characteristics as well, and will need to handle their effects accordingly when formulating our modelling strategy.

These two characteristics of a typical LGD-distribution has prompted a particular \textit{two-stage} modelling strategy in literature when estimating the LGD. As explored by \citet{leow2012LGD} and \citet{gurtler2013LGD}, this strategy involves the decomposition of the LGD into two components. In particular, the expected loss of a defaulted loan $i$ is partitioned into the probability $w_i$ of a write-off event, and into the associated loss severity $l_i$ given such a write-off event. Each component forms a separate `stage' of modelling, hence the name. The write-off risk component $w_i$ represents the probability of writing off this $i$ given a set of input variables. The loss severity component $l_i$ is the predicted loss rate of $i$ should it be written-off, also expressed as a statistical function of a (possibly different) set of input variables. Each component is then modelled separately using a chosen technique. For example, one may use binary logistic regression for $w_i$ and simple linear regression for $l_i$ across the sample $i\in\{1,\dots,N\}$ of defaulted loans. Afterwards, the estimates of $w_i$ and $l_i$ are multiplied together 
towards producing a final LGD-estimate, i.e., $\text{LGD}_i=w_il_i$.

The work of \citet{leow2012LGD} first explored a two-stage LGD-modelling approach using residential mortgage data from the UK market. However, the authors decomposed the expected LGD somewhat differently into \textit{repossession} and the associated \textit{haircut} given repossession. Both components are modelled separately using a few input variables in rendering predictions, e.g., loan-to-value (LTV), time on book, previous default indicator, and the type of security. The authors then compared this two-stage LGD-model to a single-stage LGD-model, wherein realised loss rates are directly regressed onto a set of inputs; as estimated using ordinary least squares (OLS). They found that the two-stage LGD-model faithfully reproduced the bimodal LGD-distribution, whereas the single-stage LGD-model struggled to achieve the same result. This work was extended by \citet{loterman2012benchmarking}, who compared 24 different techniques (including those used in single-stage and two-stage approaches) for estimating the LGD. The authors found that a two-stage LGD-model built with relatively simple techniques in estimating the model components (e.g., binary logistic regression for $w_i$ and linear regression for $l_i$) performed approximately the same as more sophisticated single-stage LGD-models (e.g., support vector machines and artificial neural networks). The implication is that simpler techniques fare reasonably well when tightly coupled within a two-stage modelling approach. This implication shall serve as the premise of our own work, wherein we similarly model the LGD following a two-stage approach within a broader comparative study of a few modelling techniques.

One relatively common industry-based technique for modelling the LGD is the use of \textit{run-off triangles} (ROTs). Hailing from the actuarial science literature, a ROT is type of matrix-like (two-dimensional) data structure for representing insurance claims data in aggregate. This structure typically organises aggregated losses by accident year (rows) and by the time elapsed since the accident (columns), as discussed by \citet{schmidt2006}. A ROT is generally used in modelling the emergence of aggregate losses over time by identifying longitudinal patterns in the observable "upper triangle", and projecting these patterns onto the unobserved "lower triangle" of the matrix using some technique (such as the chain-ladder method). While a ROT is used by some banks in estimating the LGD, its use is relatively understudied in literature when compared to more established LGD-modelling practices, as we shall discuss later. In particular, it is not entirely clear how such a ROT-based LGD-model would fare against a classical two-stage LGD-model, or even how to compare them exactly. As such, we intend to fill this gap with the present work.

We propose two variants of a ROT-based LGD-model to serve as our baseline. These two models are respectively estimated based on following either a single-stage approach or a two-stage approach to estimating the LGD. The two-stage ROT-based approach similarly decomposes the expected loss into a write-off risk component and a loss severity component. A composite triangle is then formed from these two components in constituting the LGD-triangle, from which LGD-estimates are then derived. We note that this particular formulation (of a two-stage ROT-based LGD-model) is itself understudied in literature, though it too suffers from the same drawbacks of any ROT, as will be discussed in the appendix.
We shall compare these baseline models against a two-stage \textit{regression-based} LGD-model, which leverages a simple modelling technique for each component of the expected LGD, estimated at the loan-level. These techniques include binary logistic regression for the write-off risk component, and a \textit{generalised linear model} (GLM) for the loss severity component. 
In total, our comparative study spans three broad LGD-modelling approaches. To the best of our knowledge, such a data-driven and in-depth comparative study has not yet been conducted, which further suggests a modicum of novelty.

As we shall demonstrate, the regression-based approach outperforms the ROT-based approach in at least two strategic ways. Firstly, the former can produce granular loan-level estimates given a rich set of input variables, whereas the latter can only render portfolio-level predictions of the LGD. This fact implies that the former can faithfully recover the empirical LGD-distribution and thereby attain a high degree of loan-level accuracy, all of which lies outside of the capabilities of a ROT-based approach. 
Secondly, the mean expected loss rate that is produced over time using the regression-based approach follows the empirical rate much closer than the one produced by the ROT-based approach. This result suggests that the former regression-based approach is more accurate than the latter, even at the portfolio-level. Greater accuracy in LGD-predictions imply that the aforementioned ECL-estimates become more realistic and "unbiased" under IFRS 9. 
Ultimately, we hope to show that a loan-level regression-based approach to estimating the LGD risk parameter is probably better than using a portfolio-level ROT-based approach.

This paper is structured as follows. We survey the rather limited literature on ROT-based LGD-models in \autoref{sec:1stROT}, followed by providing a succinct formulation of such an LGD-model. In \autoref{sec:2stage_ROT}, a two-stage variant of the ROT-based LGD-model is proposed that is based on industry practices instead of any underlying literature. 
We then present in \autoref{sec:2stage_Regression} a classical two-stage regression-based LGD-model, which serves as our challenger model to the incumbent ROT-based LGD-models. The modelling results are compared in \autoref{sec:results} across the three LGD-modelling approaches, having used a portfolio of retail credit card data from a large South African bank. This dataset spans credit cards that are observed from Jan-2013 up to Dec-2025, and contains newly-originated credit card accounts during the same sampling window.
Finally, the study is concluded in \autoref{sec:conclusion}. In the appendix, we cover: 1) a short procedure for evaluating the prediction accuracy of a ROT-based LGD-model; 2) a data-driven discussion on the demerits of a ROT-based LGD-model; 3) the selected variables in our regression-based LGD-model, and the procedure by which we selected them; 4) two time-dependent diagnostics for assessing the write-off risk model; 5) a short but novel procedure for dichotomising a probabilistic write-off risk model into binary values; and 6) the empirical LGD-distribution compared to expected variants thereof, as produced by competing regression-based LGD-models.

\section{A single-stage ROT-based approach to LGD-modelling \textcolor{red}{}}
\label{sec:1stROT}

At its core, a run-off triangle (ROT) is a two-dimensional matrix-like data structure, which is typically used in actuarial reserving for capturing insurance claims experience that is only partially observed. Casting such data into a ROT leads to a matrix that is also partially filled, i.e., it only has values in the "upper triangle". More formally, the upper triangle refers to the elements $a_{v,m}$ in an $n\times n$ matrix $A$ that satisfies $v+m\leq n+1$, where $v\geq1$ is the row index and $m\geq1$ is the column index. This matches the structural layout of an an upper anti-triangular matrix in linear algebra. Claims data are then arranged inside the upper triangle of a ROT by accident year (or cohort) across the matrix rows, and by the development period across the matrix columns. The development period $m$ can also be described as the discretely-valued time that has elapsed since the occurrence of the accident(s) in cohort $v$.

Having populated the upper triangle with claims-related data, its values can then be used in populating the missing elements in the corresponding "lower triangle"; i.e., those elements satisfying $v+m > n+1$. This operation is commonly known as "completing the ROT" (or "squaring the triangle") and is performed using some mathematical technique, which results in forecasts of the outstanding claim liabilities residing in the lower triangle. Underlying this operation is the assumption that the forecasted quantities follow a stable and systematic development pattern over the development period, as discussed by \citet{schmidt2006}. Put differently, the way in which these quantities mature over time should be consistent across cohorts so that historical data in the upper triangle can reliably inform projections in the lower triangle. This development pattern should be common across cohorts and can be viewed as a "\textit{primitive stochastic model of loss reserving}" according to \citet{schmidt2006}. In the appendix, we specifically examine this assumption (and its violation) within the context of our study and its data.

Perhaps the most common technique for completing a ROT, and hence for estimating the aforementioned development patterns, is the \textit{chain-ladder} technique. This technique is often traced back to the early work on \textit{incurred-but-not-reported} (IBNR) claim reserves, as introduced by \citet{tarbell1934ibnr} and discussed by \citet{schmidt1997}. The chain-ladder technique extrapolates historical development patterns by estimating link ratios from the data in the observed upper triangle and applying these patterns to the unobserved cells within the lower triangle.
Another pattern-estimation technique is the Bornhuetter--Ferguson (BF) approach from \citet{bornhuetter1972} for estimating IBNR losses. This BF-method combines estimated development patterns using the chain-ladder technique with an \textit{a priori} estimate of ultimate losses. 
Furthermore, the Stanard--Bühlmann method (or the "Cape Cod method"), as described by \citet{feldblum2002}, extends the BF-approach by replacing its externally-specified estimate of the expected loss with one that is data-driven. In particular, this estimate is obtained by dividing the total cumulative losses to date across all cohorts by the "\textit{total adjusted earned exposure}". The latter measures the amount of insurance coverage that has already been consumed by policyholders; i.e., it reflects the proportion of ultimate losses expected to have emerged.
\citet{verdonck2009robustification} also described a few other methods for completing a ROT, including stochastic chain-ladder models from \citet{mack1993distribution}. 
Despite these advancements in ROT-completion techniques, we shall focus our attention on the well-known and ubiquitous chain-ladder technique in the interest of simplicity.

Using a ROT in LGD-modelling seems natural since a defaulted loan also gives rise to a stream of cash flows (or recoveries) over time. Whereas insurance reserving typically projects future claim payments or incurred claims, an LGD-triangle projects the future recoveries on unresolved defaulted loans that are still actively in the workout process. The dimensions of the ROT, i.e., the accident year and development period, are respectively represented by the default cohort (or vintage) and the time since default, which is often measured monthly. Depending on the exact design of the ROT, each and every cell in the ROT can represent the aggregate recoveries at time $m$ for those defaulted loans in cohort $v$. This structure provides a portfolio-level representation of the overall development pattern of the recovery process.

However, the literature on ROT-based LGD-modelling seems to be scant. \citet{muja2015credit} estimated the portfolio-level LGD by applying the chain-ladder method on a ROT that was compiled using default data from a retail credit card portfolio. In particular, they projected the aggregated cumulative recoveries across default cohorts and the time since default. They noted that the approach is particularly suitable for sizeable retail portfolios with many small post-default recoveries. 
\citet{joubert2018default} compared ROTs with several alternative LGD-models using three South African retail credit datasets and five simulated datasets. Some of the alternative LGD-modelling techniques included linear and beta regression models, fractional response regression, and their proposed \textit{default-weighted survival analysis} (DWSA) model. They found that the latter outperformed the ROT-based method in terms of mean squared error (MSE), bias, and variance. 
Whilst they have not explicitly explored LGD-modelling, \citet{breed2021development} demonstrated the broader use of ROTs and the chain-ladder technique in the modelling of default risk term-structures under IFRS 9. The authors highlighted a few advantages of a ROT-based technique, including its simplicity, ease of implementation, and its sensitivity to segmentation schemas by which data can first be partitioned before ROT-completion. However, and as we shall demonstrate later, the success of a ROT-based technique is heavily dependent on its specification.

In our study, the LGD-specific formulation of a ROT is chiefly informed by that of \citet{alghisi2013critical}. The authors first arranged the cumulative recoveries of retail defaulted loans by default cohort and by the time since default. Then, they used the chain-ladder technique towards calculating development factors, which are then recursively applied onto the observed recoveries towards obtaining forecasts of the unobserved recoveries, thereby completing the ROT. The LGD is then derived as one minus the cumulative recovery rate. The authors stated that this ROT-based approach to LGD-modelling can leverage large unsecured retail portfolios with recurring post-default recoveries. The approach can also incorporate all available defaulted contracts, including those unresolved defaults for which the workout process is still incomplete.

It is against this background that we shall formulate our first single-stage ROT-based LGD-modelling technique. It is described as `single-stage' since it directly models the cumulative LGD rates within the triangle. Regarding the dimensions of the LGD-triangle, the $y$-axis (rows) of the triangle represents the default cohort. That is, each row represents all default spells starting in a specific calendar month $v$ within the range of months $t_1,\dots,v,\dots,t_n$, as indexed by $v$. In this context, $t_1$ signifies some starting cohort (e.g., Jan-2007), whereas $t_n$ represents the most recent cohort observed from data. The $x$-axis (columns) of the triangle refers to the time since default 
(or maturity), as indexed by $m\in\{1,\dots,o\}$, where $o$ represents the end of an imposed outcome period; e.g., $o=60$ months. Each cell within the triangle is an aggregated quantity that is calculated across all default spells within a particular cohort and maturity.

In the calculation of each cell within the triangle, first consider the total exposure at default (EAD) $E_v$ of each default cohort $v$ on the $y$-axis of the triangle. At the loan-level, let $\tau_d$ denote the integer-valued starting time of each default spell $i\in\{1,\dots,n_v\}$ within a particular $v$, where $n_v$ is the total number of default spells in $v$. Let $B_{i,t}$ denote the observed month-end balance of $i$ at each point of its spell lifetime $t\in\{\tau_d,\dots,\tau_r\}$, where $\tau_r$ is the resolution point of $i$. While both $\tau_d$ and $\tau_r$ are dependent on $i$, we shall keep the notation simple and drop $i$. Then, each $E_v$ is calculated for a given $v$ as
\begin{equation} \label{eq:EAD_sum}
    E_v = \sum_{i=1}^{n_v} B_{i,\tau_d} \, .
\end{equation}

In following the formulation of \citet{alghisi2013critical}, recoveries are treated as post-default cash flows that are discounted to the default date. Let $R_{i,t}$ represent the discounted net receipt for spell $i$ that is observed at point $t$ of its lifetime, as discounted to the default point $\tau_d$ using the applicable rates between $[\tau_d, t]$ in a chained fashion. Relevant direct workout costs are incorporated as negative cash flows within each value of $R_{i,t}$. We define the cumulative discounted receipts of cohort $v$ up to a given maturity $m$ as
.\begin{equation} \label{eq:receipts_disc}
    \bar{R}_{v,m} = \sum_{i=1}^{n_v} \sum_{t=1}^{m} R_{i,t} \, .
\end{equation}
One can then define the cumulative recovery rate as the ratio between $\bar{R}_{v,m}$ and $E_v$, i.e., the proportion of $E_v$ that has been recovered up to $m$. Let $L_{v,m}$ denote the cumulative LGD for $v$ up to $m$, where $L_{v,m}$ represents a particular cell within the LGD-triangle and is calculated as
\begin{equation} \label{eq:LGD_triangle}
    L_{v,m} = 1 - \frac{\bar{R}_{v,m}}{E_v} \, .
\end{equation}

A matrix can now be formed of dimension $n\times o$, which contains elements $L_{v,m}$ for $v\in\{t_1,\dots,t_n\}$ and $m\in\{1,\dots,o\}$. Following the ROT-framework of \citet{schmidt2006}, the sum $v+m$ corresponds to the calendar period in which each $L_{v,m}$ is calculated. Assuming a square matrix, the quantities that satisfy $v+m\leq o$ occur in the upper triangle of the matrix, and are therefore observable. Likewise, cells for which $v+m>o$ is true correspond to future calendar dates beyond the available data and constitute the empty lower triangle. We illustrate the structure of the observable upper triangle in \autoref{tab:ROT_structure}, which concludes the discussion of its construction inside the ROT.

\begin{table}[ht!]
\centering
\caption{The structure of an uncompleted run-off triangle (ROT).} 
\label{tab:ROT_structure}
\begin{tabular}{p{3cm} p{1cm} p{1cm} p{1cm} p{1cm} p{1cm} p{1cm} p{1cm}}
\toprule
\multirow{2}{*}{\textbf{Default cohort $v$}} & \multicolumn{7}{c}{\textbf{Maturity $m$}} \\
 & 1       & 2       & $\cdots$       & $m$       & $\cdots$       & $o-1$      & \textit{$o$}      \\ \cmidrule(l){1-8} 
$t_1$ & $L_{1,1}$        & $L_{1,2}$        & $\cdots$       & $L_{1,m}$          & $\cdots$       &  $L_{1,o-1}$          & $L_{1,o}$                  \\
$t_2$ & $L_{2,1}$        & $L_{2,2}$         & $\cdots$       & $L_{2,m}$          & $\cdots$       & $L_{2,o-1}$           &                   \\
$\vdots$ & $\vdots$        &  $\vdots$       &        & $\vdots$          &        &            &                   \\
$v$ & $L_{v,1}$        & $L_{v,2}$        & $\cdots$       & $L_{v,m}$          & $\cdots$       &            &                   \\
$\vdots$ & $\vdots$        & $\vdots$        &        & $\vdots$          &        &            &                   \\
$t_{n-1}$ & $L_{t_n-1,1}$        & $L_{t_n-1,2}$        &        &           &        &            &                   \\
$t_n$ & $L_{t_n,1}$       &         &        &           &        &            &                   \\ 
 \bottomrule
\end{tabular}
\end{table}

In applying the chain-ladder technique towards forecasting the lower triangle of the LGD-matrix, consider the calculation of development factors using the values in the observed upper triangle. Let $\gamma_m$ denote the development factor associated with maturity $m$, calculated from the observed cumulative LGD-values as
\begin{equation} \label{eq:dev_factor}
    \gamma_m = \frac{ \sum_v L_{v,m} }{ \sum_v L_{v,m-1} } \, .
\end{equation}
Each forecasted cumulative LGD is then computed recursively from the value in the previous column through
\begin{equation} \label{eq:forecast_rot}
    L_{v,m}
    =
    \gamma_m L_{v,m-1}
    \qquad \text{for} \quad v+m>o \, .
\end{equation}
The development factor therefore describes the proportional change in cumulative LGD between successive maturities. As the time since default increases, these development factors would generally be expected to converge towards one, as explained by \citet{alghisi2013critical}. This concludes the formulation of the single-stage ROT-based LGD-modelling technique.
\section{A two-stage ROT-based approach to LGD-modelling \textcolor{red}{}}
\label{sec:2stage_ROT}

The single-stage approach in \autoref{sec:1stROT} models the cumulative LGD directly within a single run-off triangle. An alternative, industry-informed approach is to decompose the expected LGD into two components, namely the \textit{probability of write-off} (PW) component, denoted by $P$, and the associated \textit{loss severity} (LS) component, denoted by $L'$. 
We discuss the construction of the PW- and LS-triangles in \autoref{sec:2stage_ROT_PW} and \autoref{sec:2stage_ROT_LS}, respectively. Their lower triangles are then forecast using the chain-ladder method in \autoref{sec:2stage_ROT_ChainLadder}, after which the two component triangles are combined in \autoref{sec:2stage_ROT_LGD} to form the final LGD-triangle and resulting LGD-estimates over time. This subsection also includes a brief discussion of the particular parametrisation of the ROTs used in our work.

\subsection{Constructing the probability of write-off (PW) triangle}
\label{sec:2stage_ROT_PW}

The PW-triangle retains the dimensions of the LGD-based ROT that was introduced in \autoref{sec:1stROT}. However, rather than containing cumulative LGD values $L_{v,m}$, the cells of the PW-triangle are constructed from the proportion of default balances that are subsequently written-off. Each cell within the PW-triangle signifies a particular probability of a unit of currency being written-off. We denote each cell by $P_{v,m'}$ for default cohort $v$ and overall spell age $m'$, where "spell age" is a single value that is observed across the entirety of a spell. This time definition differs slightly from that of the single-stage ROT-based LGD-model from \autoref{sec:1stROT}, where the latter time definition ("time in spell") acts like a counter in tracking the time up to the observed spell age. Put differently, a single spell is assigned to a single cell within the PW-triangle, whereas it can contribute (via its multi-period recoveries) to multiple cells within a given row of the single-stage LGD-triangle.

Default spells within cohort $v$ that were written-off at an age of $m'$ months are said to form a sample. This sample is denoted as the set $\mathcal{D}_{v,m'}$ across cohorts $v\in \{t_1, \dots, t_n\}$ and ages $m'\in \{1,\dots,o'\}$ up to a maximum (imposed) age of $o'$. We then estimate the write-off probability $P_{v,m'}$ using only the default balances $B_{i,\tau_d}$ of $i$ within $\mathcal{D}_{v,m'}$ and the total EAD $E_v$, which is expressed as
\begin{equation} 
\label{eq:PW_def}
    P_{v,m'} = \frac{ \sum_{i \, \in \, \mathcal{D}_{v,m'}} 
    B_{i,\tau_d} }{ E_v } \, ,
\end{equation}
where $E_v$ is defined as in \autoref{eq:EAD_sum}.

The $P_{v,m'}$ quantity represents the probability of a unit of currency being written-off, i.e., the proportion of the default balances in cohort $v$ with age $m'$ that were written-off. One may also estimate this $P_{v,m'}$ quantity from a volume-weighted perspective, though doing so will change the definition of $P_{v,m'}$ to being the probability of an \textit{account} being written-off in $v$ at $m'$, instead of a unit of currency. 
One compelling reason for switching to the volume-weighted variant is that the current estimator in \autoref{eq:PW_def} may very well mask heterogeneity in write-off behaviour across accounts. In particular, the estimator may become unstable if the (account-level) write-off behaviour is correlated with exposure size and/or with portfolio composition. 
There should also be consistency in the measurement level of the estimators for the other components within the overall estimate of credit risk, i.e., $PD\times LGD \times EAD$. For example, the PD-component is typically a volume-weighted estimate, which implies that write-off risk within the LGD-component ought to be similarly estimated.
However, and in the interest of comparing industry practices to our eventual regression-based LGD-model, we shall restrict our attention to the currency-weighted (or balance-weighted) variant of $P_{v,m'}$, as defined in \autoref{eq:PW_def}.

A common industry practice is to express these incremental write-off probabilities in cumulative form when calculating the development factors later. Doing so enables one to focus solely on the last column within a triangle, as a form of ``triangle aggregation'' across $m'\in\{1,\dots,o'\}$, and using the cumulative estimates at $o'$. In particular, the cumulative PW-triangle $P'_{v,m'}$ aggregates all incremental write-off probabilities up to and including $m'$, which can be expressed recursively as
\begin{equation} 
\label{eq:PW_def_cumulative}
P'_{v,m'}
=
P'_{v,m'-1}+P_{v,m'},
\quad
\text{where } P'_{v,1}=P_{v,1} \, .
\end{equation}
Thus, $P'_{v,m'}$ denotes the total proportion of default balances within cohort $v$ that were written-off with an age of up to an upper limit of $m'$. This cumulative representation implies that the PW-triangle has the same cohort--maturity structure as the cumulative LGD-triangle in \autoref{sec:1stROT}, aside from the difference in time definitions. Accordingly, \autoref{tab:ROT_structure} similarly describes the structure of the cumulative PW-triangle $P'_{v,m'}$.

\subsection{Constructing the loss severity (LS) triangle}
\label{sec:2stage_ROT_LS}


The LS-triangle is calculated using the receipts observed from default spells that were written-off or are right-censored. Regarding the dimensions of the LS-triangle, the $y$-axis again denotes the default cohort $v$, as defined previously. However, the $x$-axis (columns) now represents the \textit{time in default spell} or maturity, as measured by $m\in\{1,\dots,o\}$ up to a maximum $o$. This time definition is the same as that of the single-stage ROT-based LGD-model from \autoref{sec:1stROT}. As such, the recoveries that are observed as the spell unfolds can contribute to multiple cells within a given row of the LS-triangle.

Let $\mathcal{D}_{v,m}$ denote the set of such (written-off and right-censored) spells in cohort $v$, as observed at maturity $m$. 
Firstly, consider the total EAD $E'_{v,m}$ of all those admissible spells $i \in \mathcal{D}_{v,m}$, calculated using the corresponding default balances $B_{i,\tau_d}$ as
\begin{equation} \label{eq:EAD_sum_LS}
        E_{v,m}' = \sum_{i \, \in \, \mathcal{D}_{v,m}} B_{i,\tau_d} \, .
\end{equation}
Secondly, we re-use the notion of a discounted receipt $R_{i,t}$ from \autoref{sec:1stROT} for spell $i$ that is observed at point $t$ of its lifetime, as discounted to the default point $\tau_d$ using the applicable rates between $[\tau_d,t]$ in a chained fashion. 
For all $i \in \mathcal{D}_{v,m}$, we define the cumulative discounted receipts $\bar{R}_{v,m}'$ of cohort $v$ up to a given maturity $m$ as
\begin{equation} \label{eq:receipts_disc_2stage}
    \bar{R}_{v,m}' = \sum_{i \, \in \, \mathcal{D}_{v,m}} \sum_{t=1}^{m} R_{i,t} \, .
\end{equation}

Thirdly, let $L'_{v,m}$ denote the loss severity for cohort $v$ up to point $m$, where $L'_{v,m}$ represents a particular cell within the LS-triangle. This $L'_{v,m}$ quantity is simply calculated as
\begin{equation}
    L'_{v,m} = 1 - \frac{\bar{R}_{v,m}'}{E'_{v,m}}
\end{equation}
This calculation is analogous to the cumulative LGD $L_{v,m}$ in \autoref{eq:LGD_triangle}, except that $L'_{v,m}$ is calculated over a subset of spells (and their multi-period receipts) rather than over the full default cohort. Finally, the structure of the LS-triangle follows the same run-off triangle structure shown in \autoref{tab:ROT_structure}.

\subsection{Forecasting the PW-- and LS-triangles using the chain-ladder method}
\label{sec:2stage_ROT_ChainLadder}

Following the chain-ladder framework introduced in \autoref{sec:1stROT}, development factors are estimated separately for the PW- and LS-triangles using their respective observable upper triangles. Thus, rather than applying the chain-ladder method directly to cumulative LGD as in the single-stage approach, the two-stage approach applies it separately to the cumulative PW and LS triangles.
Furthermore, and unlike the development factors in the single-stage formulation, these factors are estimated using a rolling window of the preceding $k$ cohorts and therefore vary by both cohort and maturity (or spell age for PW).
Using the chain-ladder method, we estimate the development factor $\alpha_{v,m'}$ for cohort $v$ and age $m'$ using the corresponding elements  $P'_{v,m'}$ within the cumulative PW-triangle. This development factor is then calculated as \begin{equation} \label{eq:devFactor_PW}
    \alpha_{v,m'} = \frac{\sum_{l=1}^k P'_{v-l,m'}}{\sum_{l=1}^k P'_{v-l,m'-1} } \, ,
\end{equation}
where $k$ controls how many prior cohorts are included in the estimation of $\alpha_{v,m'}$. Likewise, the development factor $\beta_{v,m}$ for the applicable elements within the LS-triangle is calculated as
\begin{equation} \label{eq:devFactor_LS}
    \beta_{v,m} = \frac{\sum_{l=1}^k L'_{v-l,m}}{\sum_{l=1}^k L'_{v-l,m-1}}  \, . 
\end{equation}

Having calculated the development factors, consider obtaining the forecasted values within the lower part of each respective triangle (PW and LS). The crux of this operation lies in using the previous value in a given row of the lower triangle, which can be recursively specified as
\begin{align} \label{eq:PW_forecast}
    P'_{v,m'} &= \alpha_{v,m'} P'_{v,m'-1} \quad \text{for} \ v+m'>o' \, , \quad \text{and} \\
    L'_{v,m} &= \beta_{v,m} L'_{v,m-1} \quad \text{for} \ v+m>o \, . \label{eq:LS_forecast}
\end{align}
\subsection{Reconstituting the LGD from the PW-- and LS-triangles}
\label{sec:2stage_ROT_LGD}

Having separately forecasted the PW- and LS-components, the two-stage LGD is reconstituted by combining them for each default cohort and maturity. Specifically, we calculate the element-wise product of the forecasted probability of write-off $P'_{v,m'}$ and forecasted loss severity $L'_{v,m}$, using the estimates from \crefrange{eq:PW_forecast}{eq:LS_forecast}. This operation is formally defined as
\begin{equation} \label{eq:ROT_LGD}
    G_{v,m} = P'_{v,m'} L'_{v,m} \quad \text{for} \ m'=m \, .
\end{equation}
Each row in the resulting matrix corresponds to the default cohort $v$, and the columns correspond to maturities $m\in\{1,\dots,o\}$. \autoref{eq:ROT_LGD} therefore specifies the forecasted loss rate (or LGD) for every cell within the lower triangle, i.e., for $v+m>o$, thereby completing the LGD-triangle (or $G$-matrix). 
The final set of LGD-estimates over time, one for each default cohort $v$, is then taken from the last column in the $G$-matrix per industry practice, i.e., $LGD_v=G_{v,m}$ for $m=o$. This vector is typically graphed over time and blends `empirical' LGD-values with increasingly forecasted LGD-values at each time point.
To obtain a single LGD-estimate from this vector, one may simply take the average of this LGD-series, starting from the most recent $a>1$ cohorts.

To implement the ROT-based approach to LGD-modelling, several parameters must be specified beforehand. We shall use current industry practice to guide these choices for a typical credit card portfolio. Firstly, we define the length of the sampling window $t_1,\dots,v,\dots,t_n$ to span Jan-2007 up to Dec-2025. The length of this sampling window is quite wide and captures at least two externalities, i.e., the 2008 global financial crisis and the Covid-19 pandemic. Secondly, the outcome periods of both of the PW-- and LS-triangles are set to 60 months, i.e., $o=o'=60$ months. This choice allows for estimating very `granular' triangles when aggregating the PW-- and LS-quantities from loss data, though at the cost of greater data sparsity. Thirdly, the length of the rolling window used in estimating the development factors is set to $k=6$ months. This choice is relatively small compared to the outcome period, and should allow for capturing some dynamicity in the emergence of the PW and LS quantities over time. Finally, these triangles may be risk-sensitised to certain factors by segmenting the data accordingly, whereafter a separate LGD-triangle is constructed (and completed) for each segment. However, we shall not explore such segmentation in this work, purely in the interest of brevity and simplicity.

\section{A two-stage regression-based approach to LGD-modelling}
\label{sec:2stage_Regression}

At the account-level, a regression-based approach to LGD-modelling relies fundamentally on regressing historically realised loss outcomes onto a set of input variables.  As discussed by \citet{larney2025}, these realised loss outcomes can be calculated from loss data as follows.
Loan $i\in\{1,\dots,n\}$ is said to enter default at time $\tau_d$ and possibly resolve at time $\tau_r \geq \tau_d$. For each month-end spell time $t$, let $B_{i,t}$ denote the outstanding balance of loan account $i$, and let $R_{i,t}$ denote the net cash flow (receipts minus all direct costs) associated with $i$ at $t$. Note that $B_{i,t}$ is observed before recognising the effect of $R_{i,t}$.
Since the discount rate can vary over time, each cash flow should be discounted using the cumulative discount factor between the evaluation point $u$ and $t$. Let the one-period discount factor at a given time $s$ be defined as $v_{i,s}=(1+r_s)^{-1}$, where $r_s$ is the applicable nominal interest rate at a given period $s$ of loan life for $i$. Accordingly, the chain discount factor attached to the cash flow $R_{i,t}$ from time $u$ up to $t$ is expressed as
\begin{equation} \label{eq:chainedDiscFactor}
    D_i(u,t) = \prod_{s=u}^t v_{i,s} \, .
\end{equation}

\begin{figure}[ht!]
    \centering
    \includegraphics[width=0.7\linewidth, height=0.4\textheight]{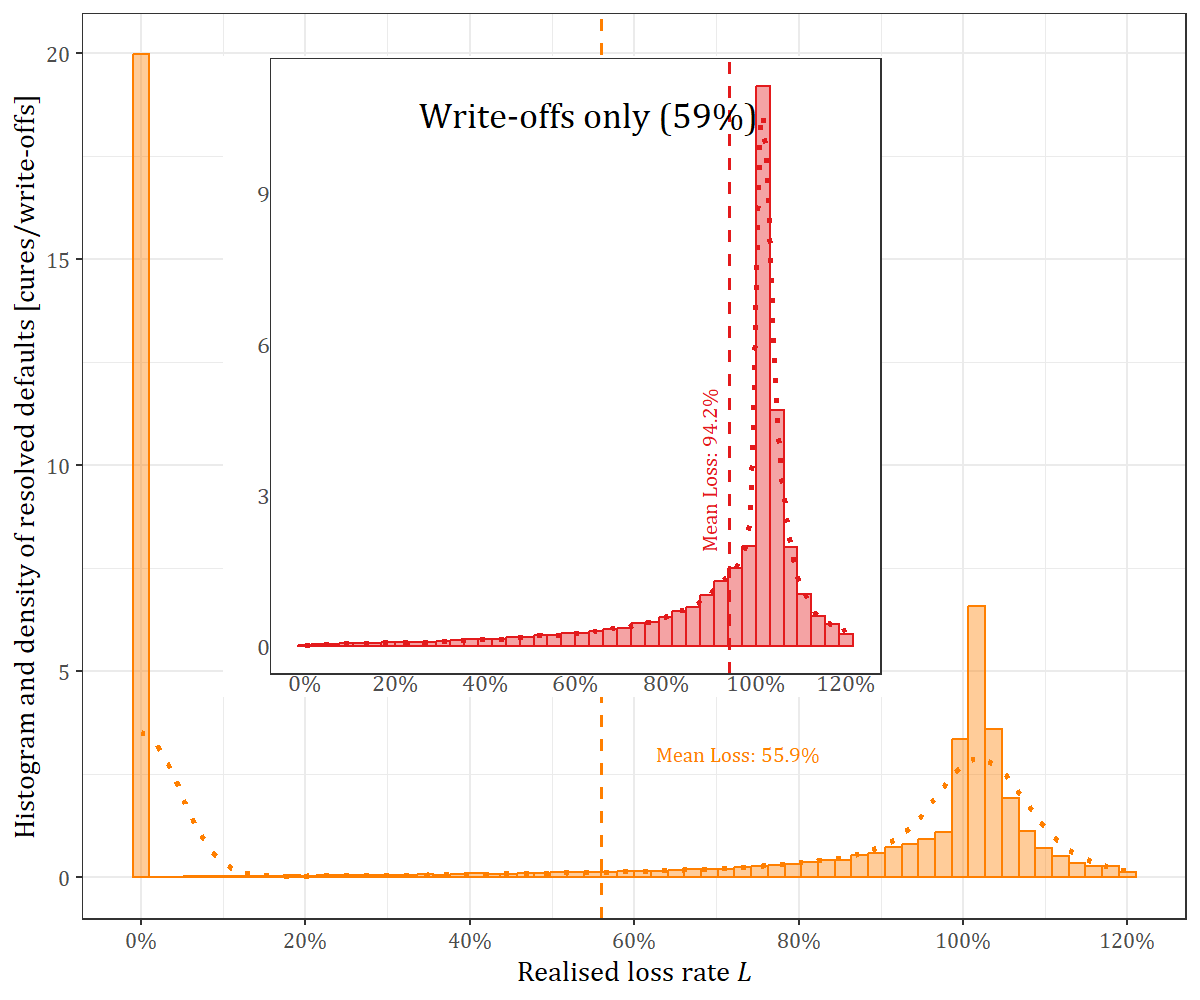}
    \caption{Distribution of empirical loss rates, as calculated at the account-level for each resolved defaulted spell using \autoref{eq:realisedLossRate}. The inset graph shows the LGD-distribution for written-off loans only. Mean loss rates are overlaid.}
    \label{fig:LGD-distribution}
\end{figure}

The realised loss rate $l_i$ is then defined as the proportion of the outstanding balance $B_{i,\tau_d}$ at the default time $\tau_d$ that is lost after accounting for all of the discounted cash flows that occur at or after $\tau_d$, up to the resolution time $\tau_r$. By definition, this $l_i$ quantity is only defined for historically written-off loans, expressed at $\tau_d$ using \autoref{eq:chainedDiscFactor} as
\begin{equation} \label{eq:realisedLossRate}
    l_i = 1- \frac{1}{B_{i,\tau_d}} \sum_{t=\tau_d}^{\tau_r} R_{i,t} D_i(\tau_d,t) \quad \text{for} \ B_{i,\tau_d} > 0 \, .
\end{equation}
The calculation in \autoref{eq:realisedLossRate} is performed for each written-off loan $i\in\{1,\dots,n_\mathrm{w}\}$ in a sub-portfolio of $n_\mathrm{w}$ write-offs. Defaulted loans that have cured are assigned a loss rate of $l_j=0$ for $j\in\{1,\dots,n_\mathrm{c}\}$ within its respective sub-portfolio of $n_\mathrm{c}$ cured loans. Note that this zero-loss-upon-cure assumption can certainly be relaxed in more advanced LGD-modelling setups in the interest of greater risk prudence.
Together, these resolved defaults (across write-offs and cures) form a sample of realised loss rates. We graph the statistical distribution thereof in \autoref{fig:LGD-distribution}, which demonstrably exhibits the two characteristics of a typical LGD-distribution: bimodality and a heavy left-skewed tail from the second mode. For written-off loans (which are about 60\% of the defaults), the mean loss rate attains a value of 94.2\%, whereas it is estimated to be 55.9\% across both cured and written-off outcomes.
Overall, the replication of the shape and features of this LGD-distribution shall become the objective of our modelling.

Across the following subsections, we shall formulate in \autoref{sec:2stage_components} the model form of each component model in the regression-based LGD-model. Then, each component model is assessed in \autoref{sec:2stage_results}, chiefly by aggregating their predictions over time and comparing these aggregates to the empirical rate. Note that additional diagnostics of each component model are provided in the appendix. In selecting variables, we follow a \textit{thematic} selection process using repeated regressions across themed subsets of input variables, as explained in the appendix. This interactive process is guided by domain expertise, model parsimony, statistical significance, goodness-of-fit, multicollinearity, and model performance on the validation dataset. Ultimately, we deem the input spaces of the two component models to be sufficiently comprehensive.

\subsection{Modelling each component in a two-stage regression-based LGD-model}
\label{sec:2stage_components}

In modelling each component (write-off risk and loss severity), we shall use the familiar modelling framework of \textit{generalised linear models} (GLMs), as discussed by \citet{agresti2015foundations}. This framework has three components that we will briefly explain as follows. Firstly, it has a \textit{random} component that specifies the response variable $Y$ and its probability distribution, with independent observations $\boldsymbol{y}=\left( y_1,\dots,y_n \right)^\mathrm{T}$ taken from a distribution in the exponential family. Secondly, the framework includes a \textit{systematic} component, which is also called a \textit{linear predictor} vector of the form $\boldsymbol{\eta}=\boldsymbol{X\beta}$, where $\boldsymbol{\beta}=(\beta_1,\dots,\beta_p)$ is an estimable $p\times 1$ coefficient vector. This coefficient vector is associated with a $p$-dimensional set of input variables $\boldsymbol{X}=\{ \boldsymbol{x}_1, \dots,\ \boldsymbol{x}_p\}$, which is an $n\times p$ model matrix that contains the $n$ observed values across the $p$ variables. 
For each observation $i\in\{1,\dots,n\}$, we let $x_{ij}$ denote the value of the $j$th input variable where $j\in \{1,\dots,p\}$, which implies the observed vector $\boldsymbol{x}_i=(x_{i1},\dots,x_{ip})$. Usually, one sets $x_{i1}=1$ in serving as the intercept term across $i\in\{1,\dots,n\}$ in the model. The linear predictor $\eta_i$ for $i$ can also be written as a linear combination of coefficients inputs, expressed as
\begin{equation} \label{eq:linearPredictor}
    \eta_i = \sum_{j=1}^p \beta_j x_{ij}, \quad i\in\{1,\dots,n\}.
\end{equation}

Thirdly, the GLM-framework has a \textit{link function} $g$ that relates the random component to the linear predictor, where $g$ is a monotonic and differentiable function. Given the conditional mean $\mu_i=\mathbb{E}(y_i|\boldsymbol{x}_i)$, this link function relates $\mu_i$ to the input variables via $g(\mu_i)=\eta_i$ for all observations $i\in\{1,\dots,n\}$, where $\eta_i$ is defined in \autoref{eq:linearPredictor}.
The unknown regression coefficients $\boldsymbol{\beta}$ are estimated via maximum likelihood. That is, values of $\boldsymbol{\beta}$ are found such that the likelihood of the observed data is maximised under the GLM specification. Equivalently, the estimation yields coefficient values for which the model-implied mean responses are the most consistent with the observed data.

In modelling the write-off risk component, which is also called the \textit{probability of write-off given default} (PWGD) model, we contend with a binary-valued random component $Y_i\in\{0,1\}$, i.e., a Bernoulli random variable. In this setup, the realisation $y_i=1$ indicates a written-off default spell $i$, whereas $y_i=0$ denotes a cured default spell. Observation $i$ is then a Bernoulli trial with probabilities $\mathbb{P}(Y_i=1)=\pi_i$ and $\mathbb{P}(Y_1=0)=1-\pi_i$. This setup is a special case of the binomial distribution with $Y_i\sim\mathrm{Bin}(n_i,\pi_i)$, where the number of trials is $n_i=1$, the success probability is $\pi_i=\mu_i$, and the observed proportion of successes is $y_i=Y_i/n_i$. 
Furthermore, we adopt the canonical link function $\log[\mu_i/(1-\mu_i)]$, which is the log odds of the response outcome 1, or the \textit{logit} of $\mu_i$. The ensuing GLM with a logit link function for each $i\in\{1,\dots,n\}$ has the form
\begin{equation} \label{eq:linkFunction_logit}
    g(\mu_i) = \ln\left( \frac{\mu_i}{1-\mu_i} \right) = \sum_{j=1}^p \beta_j x_{ij} = \eta_i \, .
\end{equation}

\citet[pp.~1--10]{hosmer2000logistic} explained that dividing $\mu_i$ with $1-\mu_i$ in \autoref{eq:linkFunction_logit} produces the \textit{odds} of the write-off event relative to the cured event, and that this odds can be transformed into a more symmetric quantity using the natural logarithm. Furthermore, the logit $g(\mu_i)$ avails a few attractive properties, e.g., its output is continuous, can range from $-\infty$ to $\infty$, and is linear in its parameters.
In this logit model, the inverse link function $g^{-1}$ is the logistic (sigmoid) function $(1+e^{-x})^{-1}$ of a given $x$, which is also why the logit model is called a \textit{binary logistic regression} model. Hence,
\begin{equation} \label{eq:linkFunction_logit_inverse}
    \pi_i = \mathbb{P}(Y_i=1 \mid \boldsymbol{x}_i) = g^{-1}(\eta_i)=\frac{1}{1+e^{_-\eta_i}} \, .
\end{equation}

In modelling the loss severity component, which is also called the \textit{loss given write-off} (LGW) model, we shall again use the GLM-framework. The random component is now continuous with $Y_i\in\mathbb{R}$, and signifies the underlying loss rate of a written-off spell $i$. We assume that the response variable follows a Gaussian distribution with constant variance $\sigma^2$, i.e., $Y_i\sim N(\mu_i,\sigma^2)$. Following experimentation, this distributional assumption yielded the best fit amongst competing choices such as the gamma distribution. 
Furthermore, an identity link function is adopted, i.e., $g(\mu_i)=\mu_i$, which implies that the conditional mean $\mu_i$ is directly modelled by the linear predictor $\eta_i$. This relationship is expressed for each $i\in\{1,\dots,n\}$ as
\begin{equation}
    g(\mu_i) = \mu_i = \sum_{j=1}^p \beta_j x_{ij} = \eta_i \, .
\end{equation}

Our model universe shall include a few variants of the PWGD and LWG-models in the interest of conducting a comparative study. These models are differentiated from one another simply by the choice of input variables in their respective input spaces; see the appendix. We formulate a \textit{basic} and an \textit{advanced} model for both the PWGD and LGW components. However, the advanced PWGD-model is split into two variants: 1) the undichotomised "Type A" PWGD-model; and 2) the dichotomised "Type B" PWGD-model. This dichotomisation step substantially improves the ability of the models to recover the empirical LGD-distribution, as will be shown later. Moreover, this dichotomisation step is based on maximising the Generalised Youden Index $J_a(c)$ across possible cut-off values $c\in[0,1]$, given a particular cost multiple $a$; and outputs the optimal cut-off $c^*$. In the appendix, we present details of this step, including a short procedure for finding $a$ optimally within our context. Ultimately, the optimal cut-off value $c^*$ is used in dichotomising the conditional write-off probability $\pi_i$ given input variables $\boldsymbol{x}_i$ of spell $i$. This operation is expressed as $\mathbb{I}(\pi_i >c^*)$, where $\mathbb{I}(\cdot)$ is the indicator function.
Our final model universe therefore includes five component models in total, where these models are used in eventually reconstituting the LGD in \autoref{sec:results}.

\subsection{Assessing each component model using resolution rates and loss rates over time}
\label{sec:2stage_results}

In examining the performance of the PWGD-models, we compile a small panel of diagnostics, though relegate most of these diagnostics to the appendix in the interest of brevity. In this section, and as our first diagnostic (of three), the aggregated predictions of the models are analysed over time (or default date); chiefly using the \textit{write-off resolution rate}. Two variants of this rate are presented, starting with the \textit{empirical} (or actual) resolution rate that is denoted by $r_y(v,\mathcal{D})$. This rate is intuitively interpreted as the proportion of spells that initially defaulted at time $v$ and that were ultimately written-off, and is constructed as follows. A given longitudinal dataset $\mathcal{D}$ is first subdivided into non-overlapping partitions $\mathcal{D}(v)$, each of which contains all default spells that started at time $v$. This $v$ indexes the $n'$ monthly time periods in $\mathcal{D}$, i.e., $t_1,\dots,v,\dots,t_{n'}$.
Given the sample of write-off realisations $y_i,i\in\{1,\dots,n\}$ where $y_i\in \{0,1\}$, we compute $r_y(v,\mathcal{D})$ at a particular $v$ as
\begin{equation} \label{eq:resolRate_emp}
    r_y(v,\mathcal{D}) = \frac{1}{ | \mathcal{D}(v) |} \sum_{i \, \in \, \mathcal{D}(v)} {y_i} \, ,
\end{equation}
where $| \mathcal{D}(v) |$ is the size of $\mathcal{D}(v)$. Right-censored spells are removed from each $\mathcal{D}(v)$ over $v$, and we call this rate the \textit{ex-censoring} resolution rate $r_y(v,\mathcal{D})$, which is subsequently used in all downstream analytics.
Ordinarily, the effects of right-censoring on the resulting time graph of $r_y(v,\mathcal{D})$ can be substantial, especially towards more recent values of $v$. The ex-censoring rate can partially mute these right-censoring effects and promote a more meaningful analysis of $r_y(v,\mathcal{D})$ over $v$.
Secondly, the \textit{expected} write-off resolution rate is calculated by averaging the predicted write-off probabilities $\pi_i, i\in\{1,\dots,n\}$ within each default cohort $v$, expressed as
\begin{equation} \label{eq:resolRate_exp}
    r_\pi(v,\mathcal{D}) = \frac{1}{ | \mathcal{D}(v) |} \sum_{i \, \in \, \mathcal{D}(v)} {\pi_i} \, .
\end{equation}

\begin{figure}[!ht]
    \centering
    \includegraphics[width=1\linewidth, height=0.7\textheight]{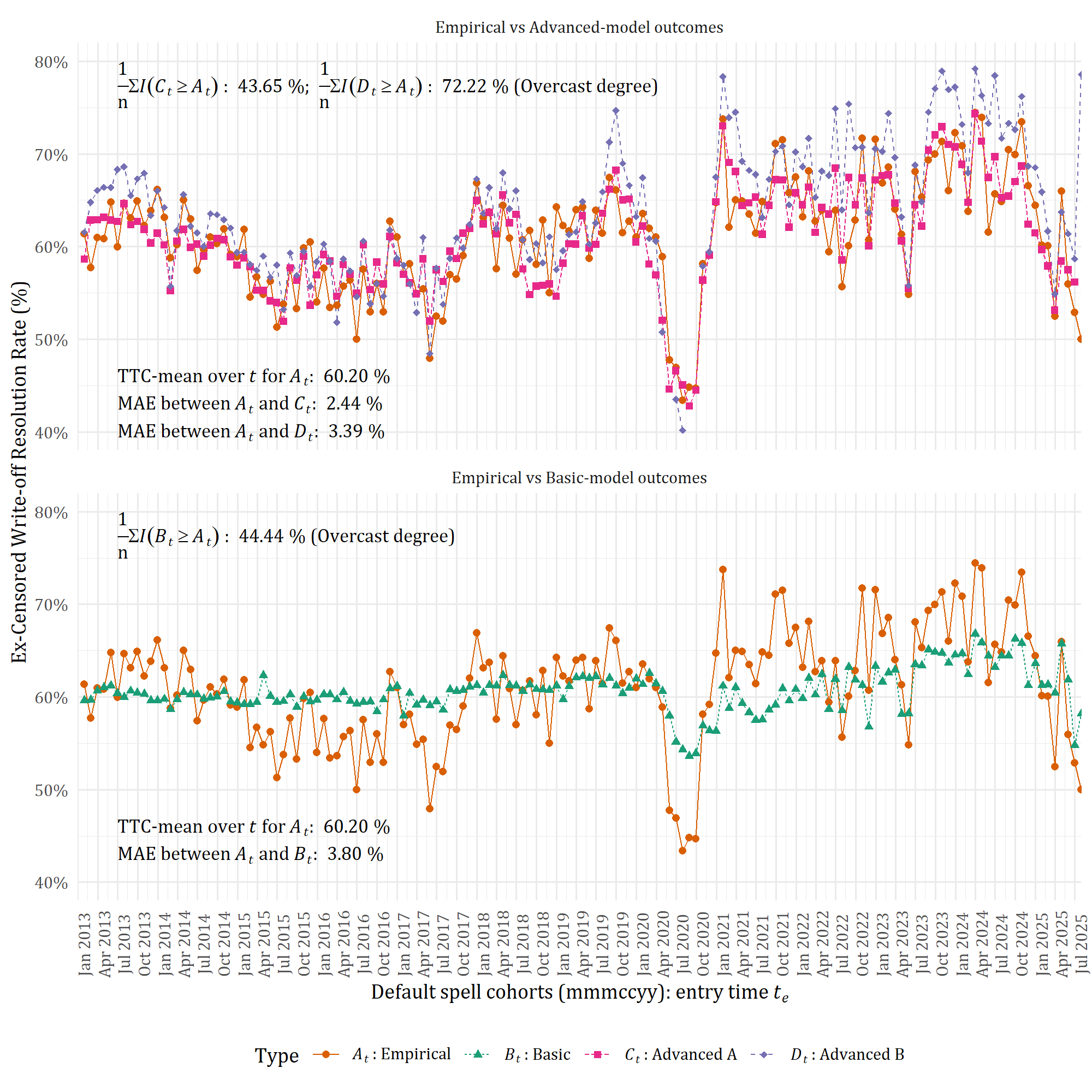}
    \caption{The expected write-off resolution rates produced by the three PWGD-models over default date, having used the full dataset $\mathcal{D}$, excluding right-censored observations. The actual (or empirical) resolution rate is overlaid across both panels. The top panel shows the expected resolution rates of the advanced Type A and B models, whilst the bottom panel shows the same for the basic model. The \textit{mean absolute errors} (MAEs) between the actual and expected rates are annotated, along with the overcast degree of each rate-pair.}
    \label{fig:ResolRate_WOff}
\end{figure}

We compare in \autoref{fig:ResolRate_WOff} the empirical and expected resolution rates over time $v$ per model . Conducting such a comparison is useful in that it can reveal periods during which patterns of structural deviation exist between the two rate types. Such a comparison also benchmarks the ability of competing models to render accurate predictions over time. A better model would produce an expected resolution rate that agrees more closely with the empirical rate.
To this point, and over most $v$, we observe that the resolution rates produced by the two advanced PWGD-models follow the empirical rate closer than that of the basic PWGD-model. 
We further quantify the level of agreement by using the MAE; lower values are better. The Type A advanced PWGD-model achieves the lowest MAE-value (2.44\%), followed by the Type B model (3.39\%), and then the basic model (3.8\%). The outperformance of the advanced models is largely attributed to their richer input space relative to the basic model. Furthermore, the aggregated outputs from both advanced models follow the empirical trend closer than that of the basic model.
Furthermore, consider the Covid-19 pandemic era of 2020; a time during which the empirical resolution rate decreased considerably, as shown in \autoref{fig:ResolRate_WOff}. This phenomenon is largely due to the bank instating large-scale payment holidays as a temporary relief measure in response to the ensuing macroeconomic shock. Clearly, the aggregated predictions of both of the advanced PWGD-models can better respond to the macroeconomic crisis, relative to those given by the basic PWGD-model. The inclusion of macroeconomic variables into the input space largely explains this outperformance during the pandemic.

Risk management principles imply that higher expected rates (or `overprediction') are preferred to lower expected rates (or `underprediction') relative to the empirical rate. This principle is quantified by desiring greater values in the \textit{overcast degree}, which is simply the proportion of times during which a given expected rate $B_t$ exceeds or equals the empirical rate $A_t$ over $t$. We express the overcast degree over $n'$ time periods as 
\begin{equation}
    \frac{1}{n'} \sum_{t=1}^{n'} {\mathbb{I}\left( B_t \leq A_t \right)}
    \nonumber \, ,
\end{equation}
which is itself annotated in \autoref{fig:ResolRate_WOff} using the expected resolution rate series from each competing model. We observe similar overcast degrees of the resolution rates from the advanced Type A and basic PWGD-models: 43.65\% vs 44.44\%. The fact that the advanced Type B model has a substantially higher overcast degree (72.22\%) is partly by design in that the underlying dichotomisation threshold $c^*$ is selected on the basis of achieving a high overcast degree. Overall, controlling for such conservatism over accuracy is an another benefit of opting for a dichotomised PWGD-model (Type B).

As a second diagnostic, we assess the discriminatory power of the PWGD-models using \textit{receiver operating characteristic} (ROC) analyses. See \citet{fawcett2006introduction} and §2 in \citet{krzanowski2009} for an overview of ROC-analysis and its summary statistic, the \textit{area under the curve} (AUC).
As calculated using the validation set $\mathcal{D}_\mathrm{V}$, the AUC-statistics of the various models are as follows. The most discriminating model is the advanced Type A PWGD-model, scoring a relatively high AUC of 82.73\%; followed by the advanced Type B model with an AUC of 74.03\%. As expected, the worst model is the basic model, attaining an AUC of only 68.93\%.
As our third diagnostic, we examine the prediction accuracy of the competing models using the Brier score (BS). This score is essentially the \textit{mean squared error} (MSE) between the binary-valued observed outcomes and the model predictions thereof; lower BS-values are better. Overall, the BS-results largely reflect the same trend as we observed in the ROC-analyses in that the best scoring model is the advanced Type A PWGD-model, with a BS-value of 0.1627. However, the second most accurate model is the basic model (0.2163) instead of the advanced Type B model (0.2441). We ascribe this result to the loss of information when dichotomising any model, which demonstrably stymies the prediction accuracy. Furthermore, we would likely have obtained even worse BS-results should we have dichotomised the basic model as well.
Lastly, both of these diagnostics (AUC-statistics and Brier scores) can also be calculated on the cohort-level and over time, as detailed in the appendix. Generally, those results echo the previous and show the outperformance of the advanced models over the basic model, as measured over time.

\begin{figure}[!ht]
    \centering
    \includegraphics[width=0.9\linewidth, height=0.4\textheight]{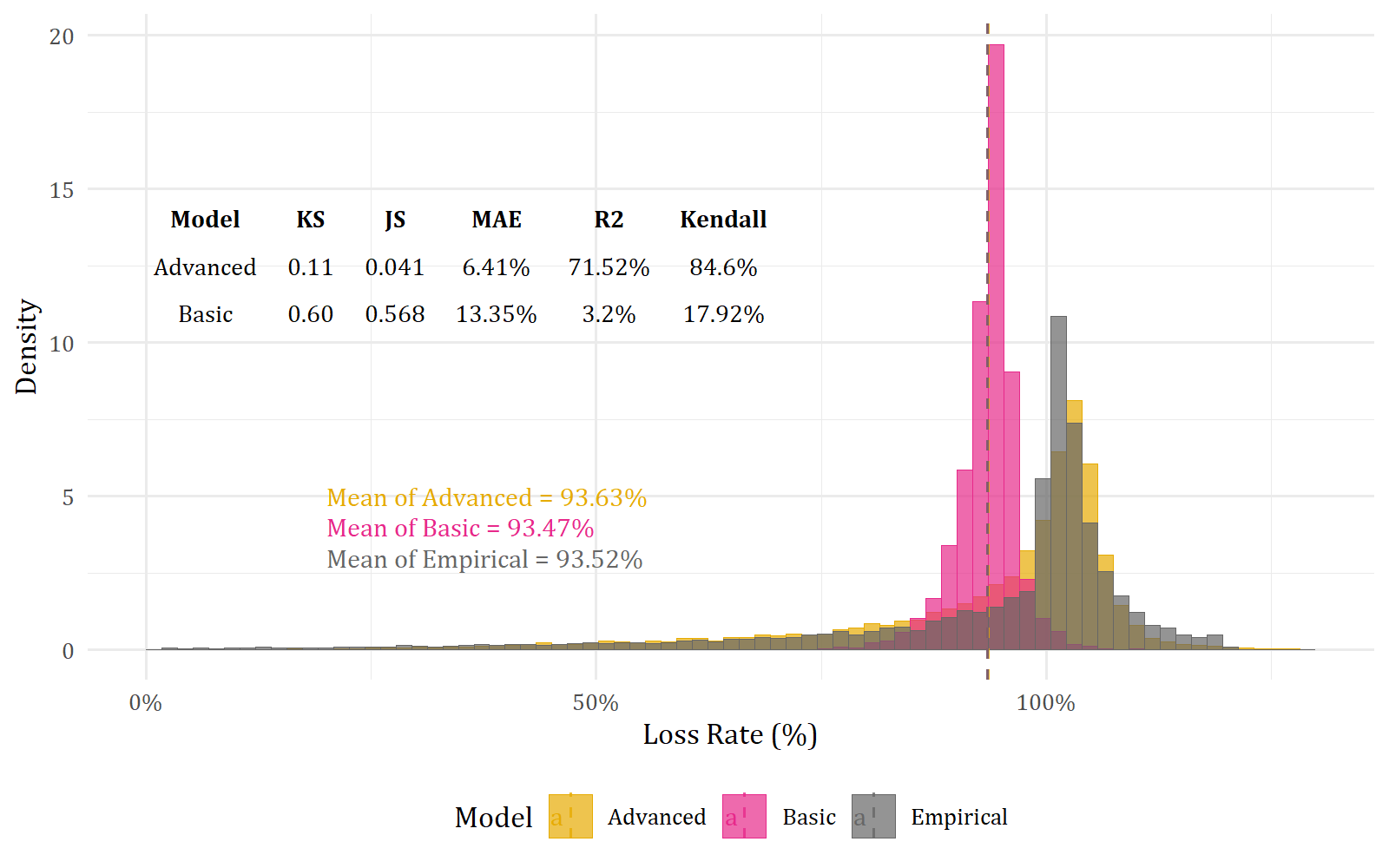}
    \caption{The distributions of expected loss rates for written-off loans, as produced by the basic and advanced LGW models using the validation set $\mathcal{D}_\mathrm{V}$. The corresponding distribution of empirical loss rates is overlaid. The mean of each distribution is annotated. Relative to the empirical loss rates, the following model performance metrics are presented: Kolmogorov-Smirnov (KS) test statistic, Jensen-Shannon (JS) divergence, MAE, coefficient of determination $R^2$, and Kendall's $\tau$ correlation coefficient.}
    \label{fig:LGW_Distr}
\end{figure}

In assessing the competing LGW-models (basic \& advanced), we compile two main diagnostics, starting with a distributional analysis. Let $l(\boldsymbol{x}_i)$ denote the loss severity prediction, as produced by a particular model, for a written-off loan $i$ with an associated set of input variables $\boldsymbol{x}_i$. The resulting distributions of the expected loss rates $l(\boldsymbol{x}_i),i\in\{1,\dots,n_W\}$ associated with $n_W$ write-offs are shown in \autoref{fig:LGW_Distr} per model, and are compared to the corresponding empirical loss distribution. Various performance metrics are calculated, including 
the Kolmogorov-Smirnov (KS) test statistic, a divergence measure (Jensen-Shannon) from \citet{Lin1991DivergenceMB}, an error measure (MAE), a correlation coefficient (Kendall's $\tau$), and the coefficient of determination ($R^2$). These metrics all indicate that the advanced LGW-model is orders of magnitude better than the basic model in fitting the empirical distribution. For example, the advanced model has an $R^2$ of 71.52\% that is close to 22 times better than that of the basic model (3.2\%). Even a visual inspection of the expected distribution of the outputs from the advanced LGW-model shows that it agrees quite closely to the empirical distribution.

\begin{figure}[!ht]
    \centering
    \includegraphics[width=0.9\linewidth, height=0.4\textheight]{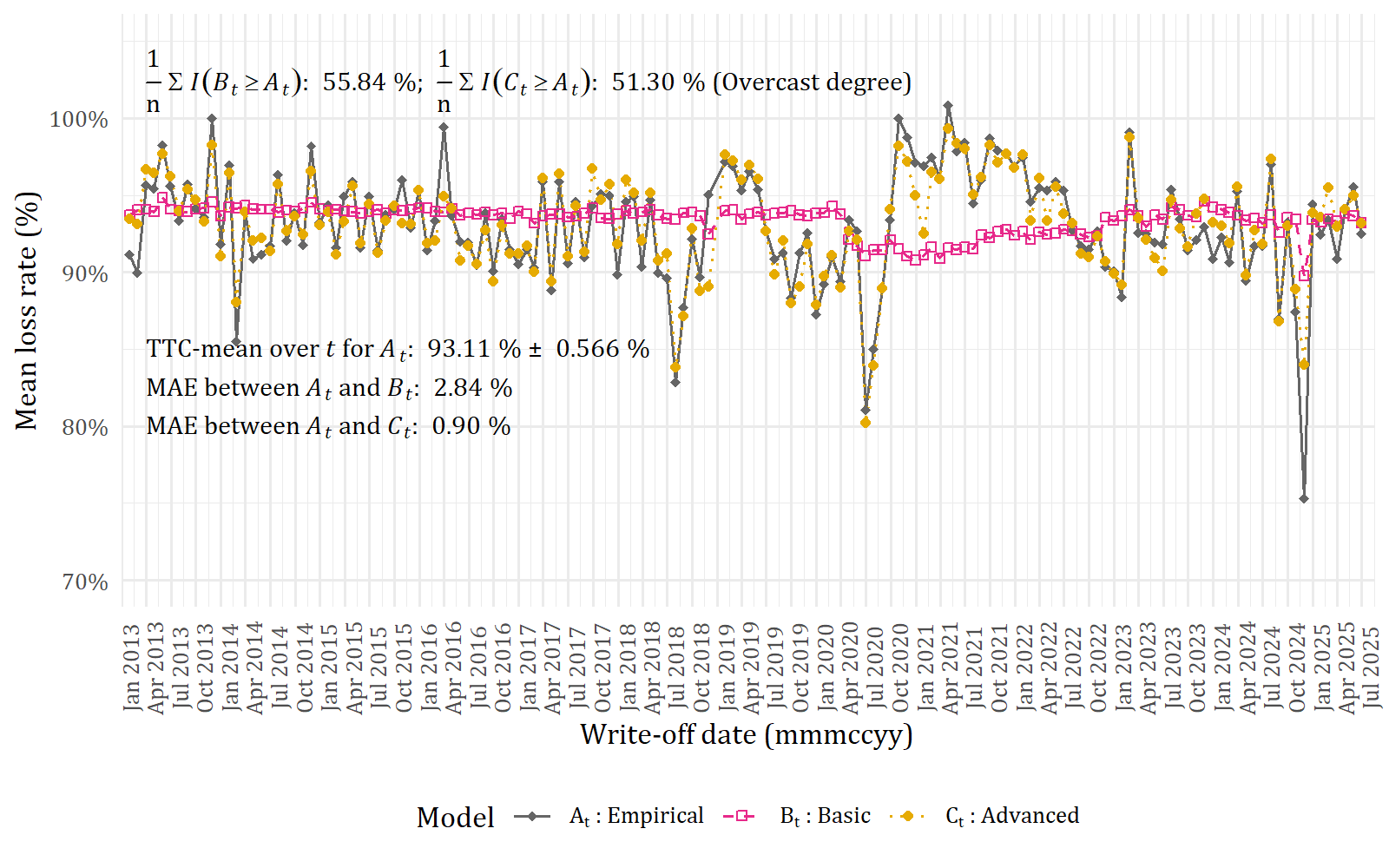}
    \caption{The mean empirical loss rate of written-off loans in the full set $\mathcal{D}$ over write-off date, compared to the mean loss rates obtained from the basic and advanced LGW-models. The MAE between the empirical loss rates and the expected loss rates are annotated per model, as is the overcast degrees between each rate pair.}
    \label{fig:LossRate_LGW}
\end{figure}

Similar to assessing the PWGD-model, we compare in \autoref{fig:LossRate_LGW} the empirical and expected mean loss rates over time per model. Consider a set of LGW-predictions $l(\boldsymbol{x}_i)$ for $i\in\{1,\dots,n\}$ that reside within a partition $\mathcal{D}(v')$ of input data $\boldsymbol{x}_i$ as observed at time $v'$, e.g., write-off date (month). The mean LGW-value at $v'$ is then simply calculated as
\begin{equation}
    \hat{l}(v')=\frac{1}{n}\sum_{i \, \in \, \mathcal{D}(v')}{ l(\boldsymbol{x}_i)} \,.
    \nonumber
\end{equation}
The empirical mean loss rate is similarly calculated, except for using the realised loss rate $l_i$ from \autoref{eq:realisedLossRate} in the summand instead of $l(\boldsymbol{x}_i)$. A better model would have closer agreement between the empirical and expected loss rates, as produced by a given model. Such agreement can again be measured using the MAE between each empirical-expected rate pair. Evidently, the advanced LGW-model achieves a lower MAE-value (0.90\%) than the basic model (2.84\%), which indicates that the former fits the data better. The mean loss rates from the basic model (shown in pink) is also relatively flat over time, which suggests that it fails in capturing time-dependent idiosyncrasies in the loss experience of write-offs.
In conclusion, all of our diagnostics show that the advanced LGW-model reliably outperforms the basic model, which is largely attributed to the richer input space of the former.

\section{Comparing the results across the LGD-modelling techniques}
\label{sec:results}

In rendering LGD-predictions, we combine the outputs from the PWGD and LGW models, respectively denoted as $w(\boldsymbol{x}_i)=\pi_i$ and $l(\boldsymbol{x}_i)$ given input variables $\boldsymbol{x}_i$ for each defaulted spell $i$. The classical two-stage LGD-model, which we shall call \textit{Type A}, is given by
\begin{equation} \label{eq:LGD_TypeA}
    L_\mathrm{A}(\boldsymbol{x}_i) = w(\boldsymbol{x}_i)\cdot l(\boldsymbol{x}_i) \, .
\end{equation}
The LGD-estimates are then reconstituted per \autoref{eq:LGD_TypeA} for both the basic and advanced model varieties. However, our experience with the Type A LGD-model is that it struggles to capture the zero-inflated nature of the empirical LGD-distribution, as shown in \autoref{fig:LGD-distribution}. Accordingly, we introduce a dichotomisation step by converting $w(\boldsymbol{x}_i)$ into a binary indicator using an optimal threshold $c^*$, as found using the Generalised Youden Index; see the appendix. This operation yields the self-styled \textit{Type B} model, which is formulated as
\begin{equation} \label{eq:LGD_TypeB}
    L_\mathrm{B}(\boldsymbol{x}_i) = \mathbb{I}\left(w(\boldsymbol{x}_i) > c^* \right) \cdot l(\boldsymbol{x}_i) \, ,
\end{equation}
where $\mathbb{I}(\cdot)$ denotes the indicator function. This structural adjustment reinstates the probability mass at zero, whilst preserving the distribution of positive losses, as produced by the LGW-model. The success hereof is demonstrated in the appendix, wherein we conduct a comparative distributional analysis across the predictions from the competing LGD-models. Overall, the Type B LGD-model is best able to recover the empirical LGD-distribution and its characteristics.

By design, the Type B LGD-model can produce zero-values in predicting the loss on cured spells. However, the Type B model is a victim of its own success since zero LGD-values imply a zero-valued provision level. Consequently, there will be no credit impairment raised against affected accounts. From a risk management perspective, this outcome is undesirable since there surely exists uncertainty in future recoveries and resolution outcomes; even on accounts expected to cure. 
One remedy to this new problem is the imposition of a positive floor onto the final LGD-value. The question then becomes how to determine such a floor, and we note that this avenue can certainly be explored in future research.
Accordingly, we have a slight preference at this stage for the Type A specification over the Type B variant of the LGD-model. However, and in the interest of providing the practitioner with options, we shall retain both specifications in the ensuing analyses.

In this section, we are chiefly interested in the evolution of the mean LGD over time, compared to the mean empirical LGD-value over time. Such an analysis can assess the degree to which cyclical variation is captured by the various LGD-models.
Consider now a set of loan-level LGD-predictions $L(\boldsymbol{x}_i), i\in\{1,\dots,n\}$ that reside within a partition $\mathcal{D}(v)$ of input data $\boldsymbol{x}_i$ observed at time $v$, e.g., default date. These predictions may come from either the Type A model $L_\mathrm{A}$ from \autoref{eq:LGD_TypeA}, or from the Type B model $L_\mathrm{B}$ from \autoref{eq:LGD_TypeB}. The mean LGD-value at $v$ is simply calculated as
\begin{equation}
    \hat{L}(v) = \frac{1}{n} \sum_{i \, \in \, \mathcal{D}(v)} {L(\boldsymbol{x}_i)}
    \nonumber \, .
\end{equation}
The corresponding mean empirical rate is similarly calculated across the loan-level empirical loss rates $l_i$ in each $\mathcal{D}(v)$, where $l_i$ is calculated using \autoref{eq:realisedLossRate} for each default spell $i$.

\begin{figure}[!ht]
    \centering
    \includegraphics[width=1\linewidth, height=0.45\textheight]{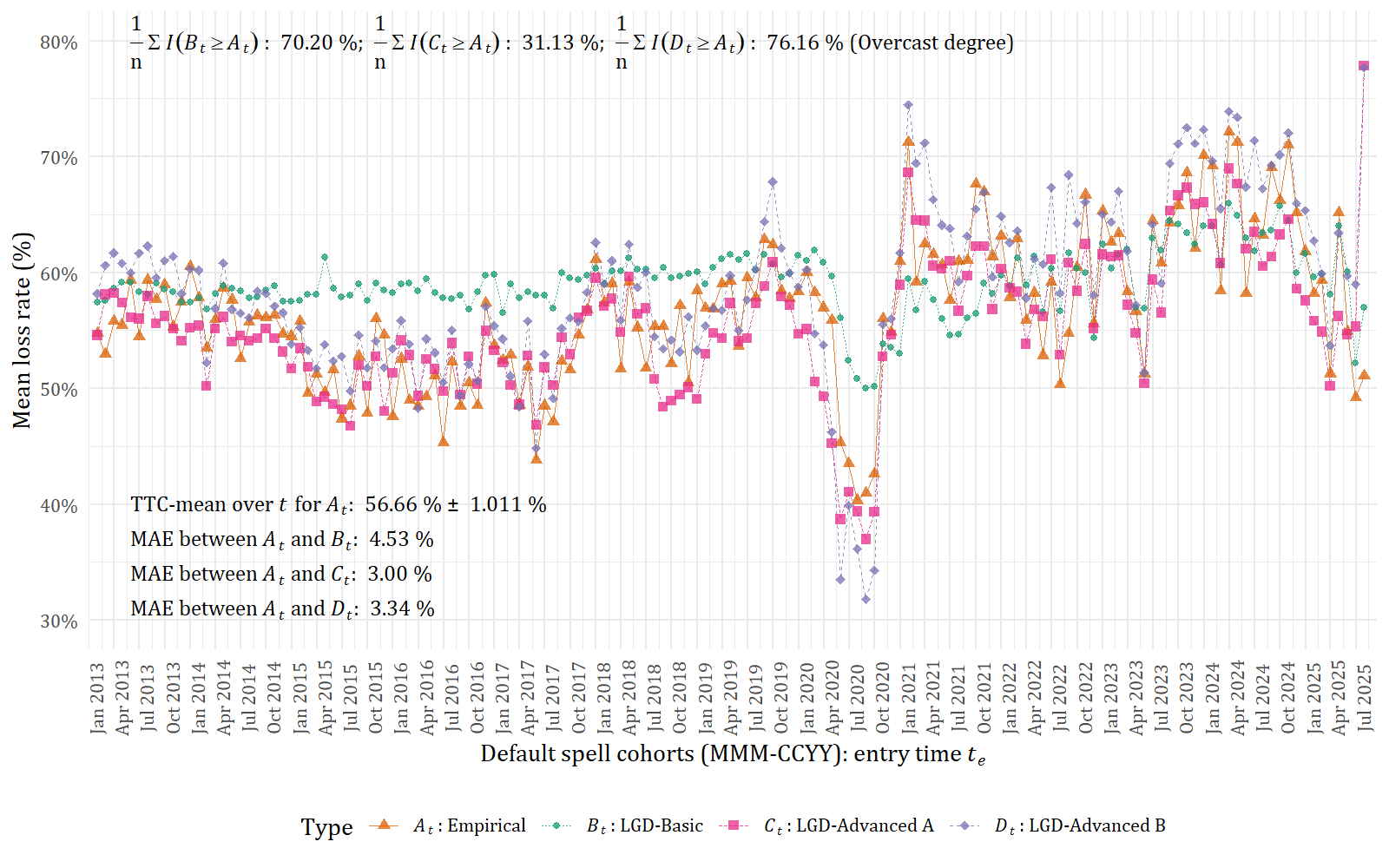}
    \caption{Mean loss rate over time using the full set $\mathcal{D}$ in comparing the empirical loss rate with that of the basic and advanced LGD-models. Overcast degrees and MAE-values are overlaid, along with the \textit{through-the-cycle} (TTC) mean empirical rate, as accompanied by a 95\% confidence interval using z-scores.}
    \label{fig:LossRate_LGD}
\end{figure}

In \autoref{fig:LossRate_LGD}, we show a time series of these mean expected LGD-values over time $v$ per model, and compare each series to the mean empirical rate over $v$. Evidently, the mean expected LGD-rates that emanate from the two advanced models (Type A + B) consistently exhibit closer agreement with the empirical loss rate over $v$, than what is the case with the basic model. We believe that this result is largely due to the richer input space of the advanced LGD-models. This outperformance is further evidenced by the substantially lower MAE-values of 3\% and 3.34\% for the two advanced models, compared with the 4.53\% MAE-value for the basic model. 
Regarding the Covid-19 pandemic era, we attribute the associated decline in the mean empirical rate to the extensive payment relief/holiday programmes that were implemented at the time. These interventions allotted additional time for distressed accounts to cure from default, thereby alleviating write-off risk and reducing the mean empirical rate. Ultimately, the two advanced models react demonstrably better to the pandemic than the basic model.

The LGD-models differentiate themselves further by their innate degree of `conservatism', as measured using the overcast degree from \autoref{sec:2stage_results}. In particular, the basic model produces a relatively smooth series of aggregated predictions over time $v$, resulting in a muted response to cyclical movements whilst overstating the empirical loss rate in general. This phenomenon manifests in a relatively large overcast degree (70.2\%) for the basic model.
In contrast, the advanced Type A model is substantially less conservative with an overcast degree of only 31.13\%, even though it is the most accurate model. Like the basic model, the advanced Type B model is considerably more conservative than the Type A variant, and over-casts the mean empirical rate in 76.16\% of periods. While this behaviour is desirable from a risk management perspective, it is achieved at the expense of a modest reduction in predictive accuracy; itself measured here using the MAE.

\begin{figure}[!ht]
    \centering
    \includegraphics[width=0.8\linewidth, height=0.4\textheight]{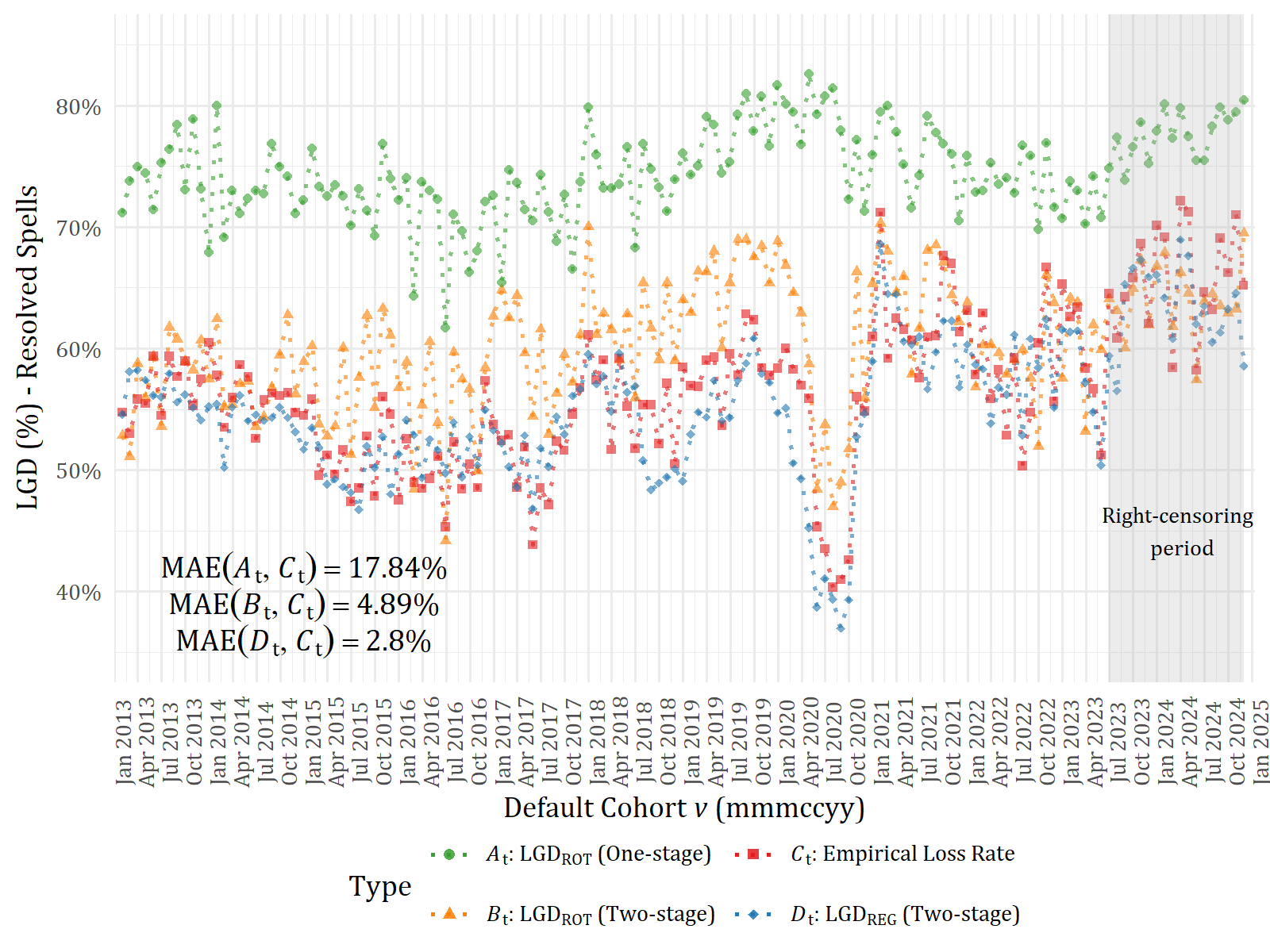}
    \caption{Comparison of the empirical loss rate over time to its expected LGD-counterparts. These counterparts include the two ROT-based LGD-values (one and two-stage), and those aggregated predictions produced by the chosen two-stage regression-based LGD-model (Type A). The mean values of each series are overlaid, as are the MAE-values between each series. The grey box indicates the period during which right-censoring is material.}
    \label{fig:LossRate_ROTvsReg}
\end{figure}

In \autoref{fig:LossRate_ROTvsReg}, we again compare the mean empirical loss rate to its expected (or predicted) counterparts over time. Two broad groups of expected LGD-rates exist: 1) those arising from the single-stage and two-stage ROT-based LGD-modelling approaches; and 2) those emanating from the chosen regression-based approach (Type A).
While the expected LGD-values from both groups broadly capture the long-run level and (slightly upwards-moving) direction of realised losses, the regression-based approach produces values that align much closer to the empirical rates. The divergence between empirical and expected rates is particularly pronounced for the single-stage ROT-based approach. Moreover, consider the upward trend in the empirical loss rate between Jul-2015 and Jul-2017. In comparing this trajectory to those observed in the expected LGD-estimates from the two ROT-based approaches, we note the following. The expected LGD-estimates have a significantly delayed reaction to the empirical trend.
In contrast, the regression-based model produces rates that closely follow the overall empirical trends, suggesting a greater ability to adapt to changing portfolio and economic conditions. This difference in rates stems directly from the differences between the underlying modelling methodologies, as will be summarised shortly.

The ROT-based approach relies primarily on historical recovery experience and therefore reacts relatively slowly to emerging changes, depending on the ROT-specification; as discussed in the appendix. In contrast, the regression-based framework incorporates various input variables (e.g., loan-level and macroeconomic covariates) that allow the expected LGD-values to evolve rapidly with prevailing conditions. The regression-based approach also avails LGD-estimates \textit{at the loan-level}, which better aligns with the spirit of the IFRS 9 accounting framework. Furthermore, the ROT-based approach demonstrably provides conservative estimates, producing higher LGD-values than the empirical loss rates. While a degree of conservatism may be desirable from a risk management perspective, excessive conservatism can lead to unnecessarily elevated levels of loan impairment provisions. In turn, such excessive conservatism can stymie the efficient allocation of capital in funding new loans or other business operations.
\section{Conclusion}
\label{sec:conclusion}

In credit risk modelling, the LGD risk parameter is commonly modelled by following a two-stage loan-level modelling approach. Each stage corresponds to a separate
component model (write-off risk and the loss severity), and the product thereof constitutes the expected loss. The rationale of such a modelling strategy is that it can faithfully recover the typical shape of the empirical LGD-distribution. This LGD-distribution is commonly characterised by bimodality and a heavy skew in its tails, owing to the dominance of zero-valued cures.
However, some banks employ an approach based on a \textit{run-off triangle} (ROT) from the actuarial science literature in modelling the LGD at the portfolio-level. This modelling strategy is a relatively understudied topic within the context of credit risk modelling. It is certainly true that some synergies exist between insurance loss data and credit loss data that would seemingly support following this ROT-based approach. However, the ROT-based approach also has a slew of demerits, as we have discussed in the appendix. These demerits make the approach intractable when compared to the more established \textit{regression-based} LGD-modelling approach that is trained at the loan-level. Perhaps the greatest of these demerits is the inability of a ROT-based LGD-modelling approach to produce loan-level LGD-estimates. The natural question then arises: which approach is `best' in producing the most credible and accurate LGD-estimates given empirical loss data?
It is in this context that we conduct an in-depth comparative study between a loan-level LGD-modelling strategy and a ROT-based one towards filling the obvious gap in literature.

In our study, we formulated two variants of a ROT-based LGD-model that are largely based on industry experience. These variants include a single-stage model and a more sophisticated two-stage model, which decomposes the expected loss into a write-off risk component and a loss severity component (or triangles). Both of these ROT-based LGD-models are then compared against a classical two-stage regression-based LGD-model. The two components within this regression-based LGD-model are estimated using simple modelling techniques, which includes binary logistic regression for write-off risk, and a Gaussian GLM for the loss severity. We purposefully selected these techniques for their simplicity and ubiquity in banking, without reaching for more sophisticated modelling techniques such as those from the machine learning literature.
In so doing, we believe that our comparative study is fairer to the simple ROT-based approach, without compromising our subsequent findings.

We fitted the two variants of the ROT-based LGD-modelling approach to a portfolio of credit card data from a large South African bank, and note the following main demerits and general findings. 
The aggregation logic of the ROT-based approach is sensitive to extreme observations, which demonstrably affects the downstream LGD-estimates. We know from literature that the ROT-based approach is also prone to becoming over-parametrised relative to the available data, which is further compounded when segmenting the data prior to estimating a ROT within each segment. Furthermore, the recursive estimation of development factors (and therefore subsequent forecasts) are prone to error propagation whereby bias in earlier cell-estimates can be carried forward into forecasting each new ROT-cell. In reality, these development factors are also not necessarily stable across default cohorts, which violates a key assumption of the ROT-based approach.
Lastly, compiling LGD-values over time (default date) when following this approach amounts to blending the concepts of `empirical' and `expected' quantities, especially as one moves to more recent default cohorts. This operation complicates model assessment, which can easily hoodwink the practitioner into thinking that the completed ROT fits the loss experience well. We found otherwise, having explicitly compared the `forecast' time series of ROT-based LGD-values against the mean empirical loss rate over time. Specifically, we found substantial divergence between these two rate pairs, which further brings into question the overall tractability of the ROT-based approach to LGD-modelling; as critically discussed in the appendix.

Using the same data, we developed a regression-based LGD-model that consists of two component models: the \textit{probability of write-off given default} (PWGD) model, and the \textit{loss given write-off} (LGW) model.
Both component models are assessed individually, and our assessment spans a number of diagnostics and modelling aspects, e.g., discriminatory power (ROC-analysis), predictive accuracy (Brier scores), and distributional analyses. Accordingly, each component model fits the data well and produces credible and sufficiently accurate predictions of the loss experience (or elements thereof).
Having built our regression-based LGD-model, we finally compared the aggregated LGD-estimates from each approach to the mean empirical rate over time. Our results indicate that both ROT-based approaches exhibit significant divergence from the empirical rate. In contrast, the regression-based approach produces aggregates that closely follow the time-sensitive trends in the empirical rate, which clearly underscores greater prediction accuracy. This result also suggests that the underlying regression-based LGD-model has a greater ability to adapt to changing portfolio and economic conditions, relative to the ROT-based LGD-model(s).
Therefore, our work demonstrably showcases the benefits of a loan-level modelling strategy, even when using relatively simple modelling techniques in estimating each component. By implication, 
a loan-level regression-based approach to estimating the LGD risk parameter is probably better than using a portfolio-level ROT-based approach.

Future work can explore other techniques for completing a ROT, and then repeat the comparative study. Whilst we have opted to use the simple and ubiquitous chain-ladder technique in completing a ROT, other techniques (e.g., the Bornhuetter-Ferguson technique) may produce different LGD-estimates over time.
While data availability is typically problematic, the study could also be reperformed on other credit datasets and/or on portfolios other than credit cards. Doing so can help generalise our findings. Furthermore, our regression-based LGD-model serves as a foundational basis on which future research can improve. It is certainly possible to replace the modelling technique that underpins the PWGD and LGW models with more sophisticated techniques, whereafter the comparative study can be reperformed. Doing so would likely engender greater predictive accuracy in the LGD-estimates, thereby strengthening the case for moving away from ROT-based LGD-models.


\appendix
\section{Appendix}
\label{app:appendix}

In \autoref{app:ROT_predictionAcc}, we formulate and demonstrate a short procedure for evaluating the out-of-sample predictive performance of a ROT-based LGD-model. Then, the merits and demerits of a ROT-based LGD-model are discussed in \autoref{app:ROT_demerits}. We present and describe in \autoref{app:InputSpaces} the selected input variables respective to the write-off risk model and the loss severity model. This discussion includes the methodology by which these variables are selected. In \autoref{app:Diagnostics_2Stage_Regression}, two ancillary time-dependent diagnostics are provided for assessing the aforementioned write-off risk model. Then, we present and demonstrate a short procedure in \autoref{app:PWGD_Dichotomisation} for dichotomising a probabilistic classifier (i.e., the PWGD-model) using the Generalised Youden Index. Lastly, a distributional analysis is provided in \autoref{app:LGD_Distributions} in comparing the empirical LGD-distribution to a small range of expected counterparts, as produced by competing regression-based LGD-models.

\subsection{A procedure for evaluating the loan-level prediction accuracy of a ROT-based LGD-model}
\label{app:ROT_predictionAcc}

To evaluate the predictive performance of a ROT-based LGD-model, the entire sampling window of default cohorts can be split into two non-overlapping datasets. The first set is used in estimating (or `training') the ROT-based model, i.e., aggregating each cell within the upper triangle. The second (out-of-time) set can be used in evaluating the predictions of the ROT-based model (i.e., the cells in the lower triangle) during the out-of-time period. The training set therefore has a contracted sampling window during which only the default cohorts $t_1, \dots, t_v, \dots,t_{n'}$ are extracted up to a certain point $t_{n'}$. In contrast, the validation set has the full history of cohorts $t_1,\dots,t_{n'},\dots,t_n$ available for $n'<n$. This sampling schema enables the evaluation of the predicted LGD-values within the lower LGD-triangle for those cohorts during the interval $[t_{n'},t_n]$, for which we have realised LGD-values available.

For each cohort-maturity cell in the lower triangle, or the tuple $(v,m)$, the prediction accuracy of the ROT-based LGD-model is evaluated by comparing the subsequent LGD-estimate $G_{v,m}$ to the realised loss rates across all spells in that cell. Let $\mathcal{J}_{v,m}$ denote the set of resolved defaults within cell $(v,m)$ and define the corresponding cell size as $n_{v,m}=\mid \mathcal{J}_{v,m} \mid$. 
Consider that the ROT-based LGD-modelling approach can only produce a single LGD forecast $G_{v,m}$ per cell. In contrast, each cell $(v,m)$ contains the realised loss rates of all resolved accounts $i \in \mathcal{J}_{v,m}$, which forms an empirical loss distribution for each $(v,m)$. We may compare any particular LGD-prediction $G_{v,m}$ for $v+m>o$ with these realised loss rates within each $(v,m)$. Then, consider the \textit{mean absolute error} (MAE) to be an appropriate accuracy measure, which can be computed between $G_{v,m}$ and the associated realised loss rates. Accordingly, the cell-level MAE is defined within the lower LGD-triangle as
\begin{equation} \label{eq:MAE_rot_LGD}
    \textrm{MAE}_{v,m} = \frac{1}{n_{v,m}} \sum_{i \, \in \, \mathcal{J}_{v,m}} \left| l_i- G_{v,m} \right| \quad \text{for} \ v+m>o\, .
\end{equation}
This calculation yields a triangle of MAE-values with the same dimensions and structure as that of the lower LGD-triangle. This collection of MAEs provides a view of the forecasting performance across cohorts and maturities. A \textit{grand} MAE can then be calculated as the average of these MAEs, which yields an overall measure of prediction error.

\begin{figure}[ht!]
    \centering
    \includegraphics[width=0.9\linewidth, height=0.5\textheight]{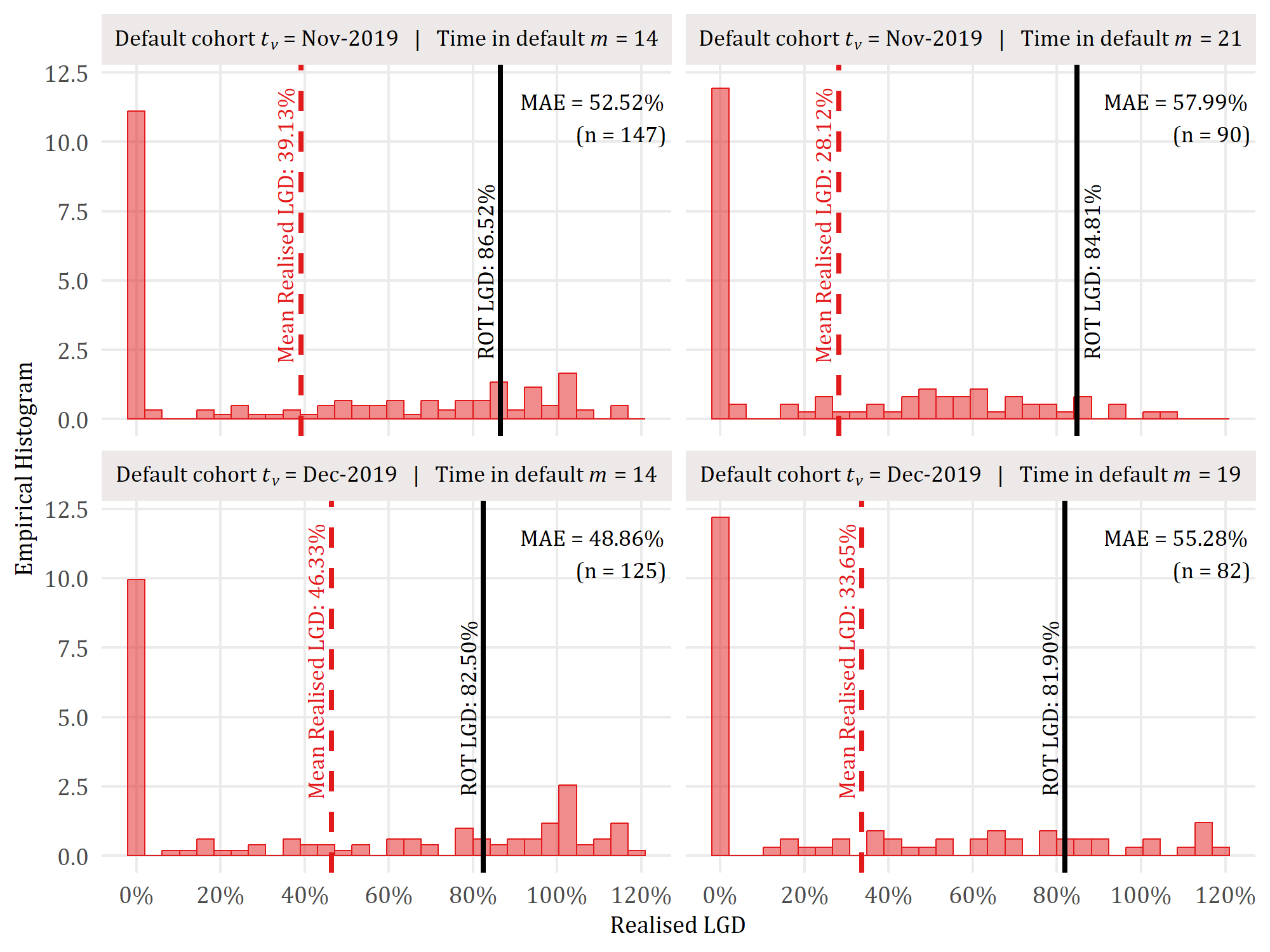}
    \caption{Empirical distributions of realised loss rates for selected cohort-maturity cells $(v,m)$ within the LGD-triangle, as shown across the panels. The ROT-based LGD-estimate $G_{v,m}$ is overlaid as a solid black line, along with the mean realised loss rate in red. The MAE between these tow lines is overlaid, together with sample sizes $n$.}
    \label{fig:ROT_LGDs}
\end{figure}

Accordingly, we use \autoref{eq:MAE_rot_LGD} to evaluate the out-of-sample accuracy of the LGD-prediction $G_{v,m}$ that emanates from the ROT-based LGD-model for any cohort-maturity cell $(v,m)$ in the lower LGD triangle.
\autoref{fig:ROT_LGDs} presents histograms of the realised loss rates of resolved default spells, which are categorised across four particular cells $(v,m)$ that were chosen using expert judgment. For each cell, the red dotted line denotes the empirical mean LGD across the spells in the cell, whereas the black solid line represents $G_{v,m}$. Evidently, the $G_{v,m}$ values exceed the empirical means in every one of these four example cases. Furthermore, the MAE-values calculated using \autoref{eq:MAE_rot_LGD} range between 49\% and 58\%, which indicates substantial prediction error. We note that the data is extremely sparse in certain cells, which attests of the over-parametrisation problem of ROTs, as will be discussed later in \autoref{app:ROT_demerits}. Nonetheless, the evidence suggests that the ROT-based LGD-modelling approach performs poorly in capturing the central tendency of realised losses, ultimately overstating LGD-estimates.

\begin{figure}[ht!]
    \centering
    \includegraphics[width=0.6\linewidth, height=0.35\textheight]{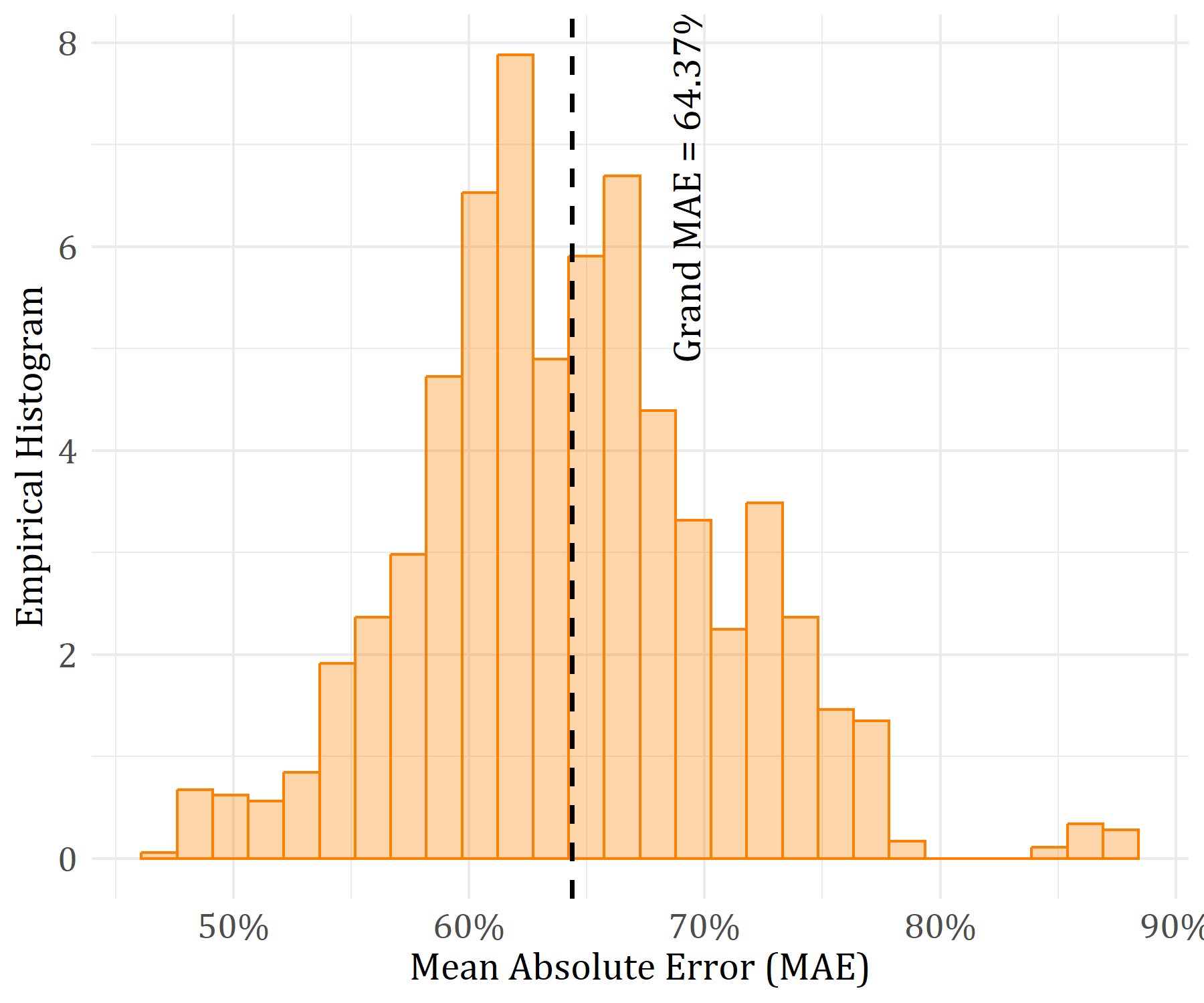}
    \caption{Distribution of cell-level MAE-values for the ROT-based LGD forecasts across all cohort-maturity pairs $(v,m)$ in the lower LGD-triangle. The grand MAE is overlaid.}
    \label{fig:ROT_GrandMAE}
\end{figure}

Beyond the four illustrative cells in \autoref{fig:ROT_LGDs}, we also consider the full distribution of these cell-level forecasting errors $\mathrm{MAE}_{v,m}$ across all cohort-maturity pairs $(v,m)$ in the lower triangle, as shown in \autoref{fig:ROT_GrandMAE}. The grand MAE of 64.37\% suggests that the ROT-based LGD-estimates $G_{v,m}$ systematically fail to approximate the realised loss rates across all $(v,m)$, with substantial average errors relative to the true losses. All together, these findings outline the inability of the ROT-based LGD-modelling approach to predict the empirical LGD-values with sufficient accuracy at a more granular level.

\subsection{A critical review of a ROT-based LGD-model}
\label{app:ROT_demerits}

We discuss in \autoref{app:ROT_demerits_nonrobust} the statistical non-robustness of the base aggregation logic within the construction of a ROT, relying on robust-statistics theory. Then, the inherent over-parametrisation of ROTs relative to the available data is highlighted in \autoref{app:ROT_demerits_overParams}. We outline in \autoref{app:ROT_demerits_recursive} the problem of error propagation when using a ROT to forecast. Finally, we compare in \autoref{app:ROT_demerits_divergence} the two-stage ROT-based LGD-estimates with the mean empirical loss rate over time and draw some conclusions.

\subsubsection{Statistical non-robustness of the aggregation method within a ROT-model}
\label{app:ROT_demerits_nonrobust}

Regarding the PW-triangle from \autoref{sec:2stage_ROT_PW}, each set $\mathcal{D}_{v,m'}$ contains information on the EAD-values of accounts in the default cohort $v$ that have been written-off at an age of $m'$ months. The base aggregation logic of creating such a triangle is to sum these EAD-values together in each cell. A key modelling assumption of the ROT-based approach is that the distribution of these EAD-values within each $\mathcal{D}_{v,m'}$ is reasonably symmetric. In particular, when the aggregation (in constructing a ROT) relies on an estimator that is sensitive to extreme observations, then the validity of the resulting point estimate depends critically on the underlying distribution being well-behaved within each $\mathcal{D}_{v,m'}$.
This means that a small subset of unusually large/small EAD-values should not dominate the distribution, lest the eventual sum of EAD-values misbehaves as a summary statistic in the presence of such outliers.

Robust-statistics theory shows that the arithmetic mean is nowhere robust, which means that even a single extreme observation can force the estimator arbitrarily far from its true value. \citet{hamepl1971Robustness} formalised this insight by introducing the concept of the \textit{breakdown point}, which is defined as the smallest number of replaced observations that causes an estimator to assume an arbitrarily large (or small) value. This concept is also called the \textit{finite breakdown point} when applied to a fixed sample size $n$. Taking the limit as $n\rightarrow\infty$ yields the \textit{asymptotic breakdown point}, which lies between 0 and 0.5. A value of 0 corresponds to complete non-robustness, whereas 0.5 is the theoretical maximum, as achieved by highly robust estimators such as the median. For the arithmetic mean, the finite sample breakdown point is $1/n$, since altering just one data point is enough to move the mean by an arbitrarily large amount. As the sample size increases, this breakdown point converges to zero, giving the mean an asymptotic breakdown point of 0. By implication, the arithmetic mean is non-robust and is highly sensitive to outliers.

This concept of breakdown points is also applicable to the ROT-based LGD-modelling approach since it calculates each PW-cell $P_{v,m'}$ from \autoref{eq:PW_def} using the \textit{sum} of the EAD-values within each set $\mathcal{D}_{v,m'}$. As the sum is simply a scaled version of the mean $\bar{x}$, i.e., $\sum_i x_i=n\bar{x}$, it inherits the same robustness properties of the mean. Consequently, the sum has the same finite breakdown point of $1/n$ and an asymptotic breakdown point of 0. These properties imply that any single extreme EAD-observation within a $\mathcal{D}_{v,m'}$ can disproportionately distort the aggregated value across the spells $i\in\mathcal{D}_{v,m'}$. In practice, one can assess whether this modelling assumption is reasonable by inspecting the empirical distribution of EAD-values within each $\mathcal{D}_{v,m'}$. Indicators such as a high skewness statistic, the presence of extreme values, or long right tails can suggest that the sum is likely to be unduly influenced by a small number of atypical EAD-values.

\begin{figure}[ht!]
    \centering
    \includegraphics[width=0.9\linewidth, height=0.5\textheight]{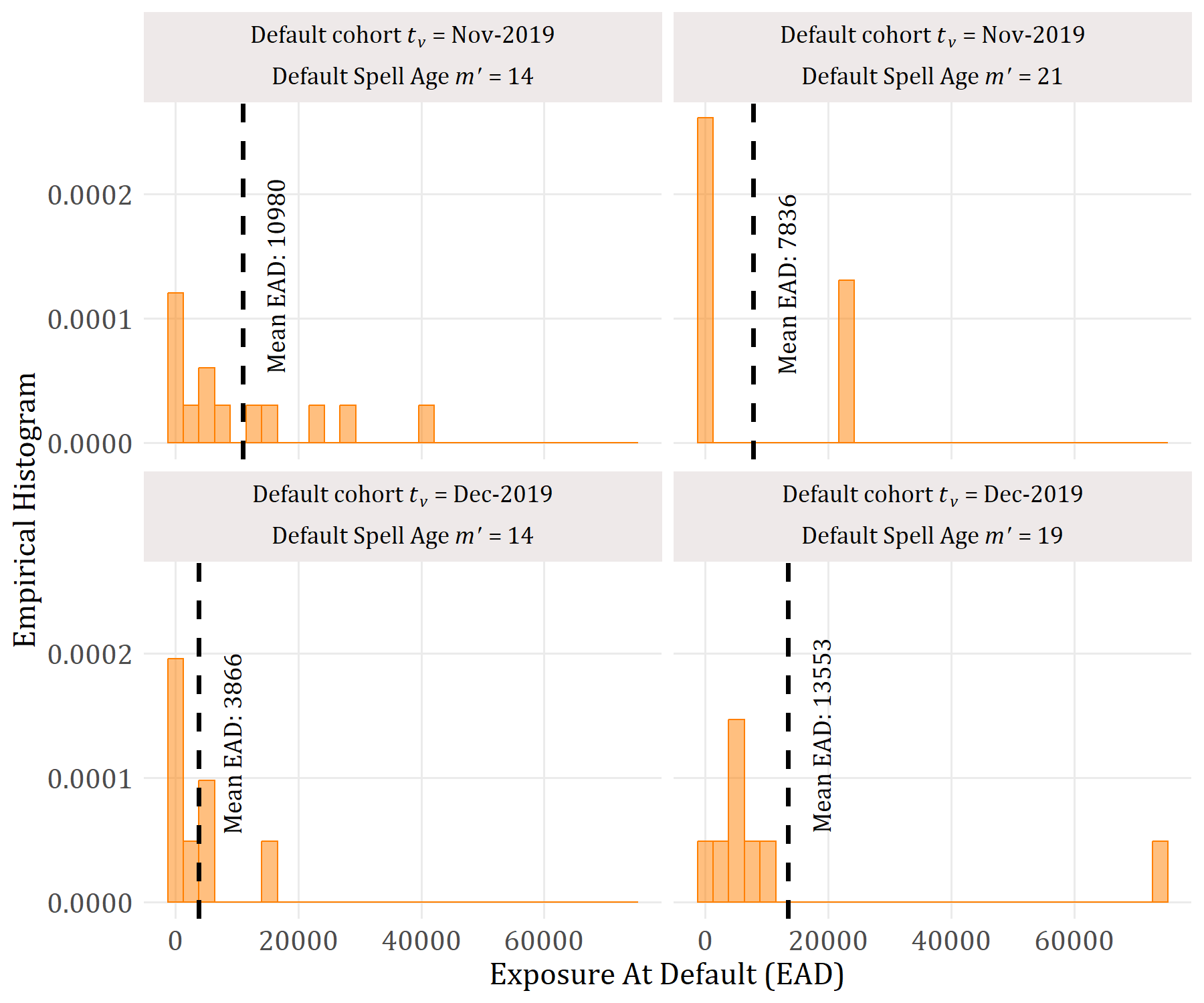}
    \caption{Empirical distributions of Exposure At Default (EAD) values for selected cohort-maturity cells $(v,m')$ within the PW-triangle, as shown across the panels. The mean EAD is overlaid in each case.}
    \label{fig:ROT_EADs}
\end{figure}

The empirical EAD-distributions in \autoref{fig:ROT_EADs} highlight the concern of statistical non-robustness. At smaller default spell ages, the EAD-values have a more pronounced right-skewed distribution. The sample size decreases substantially for larger spell ages, and the effect of outliers become even more prominent. These distributional features represent clear departures from the assumed symmetry of the EAD-distributions across partitions when following the ROT-based approach to LGD-modelling. In turn, these departures create conditions under which the robustness of the cell estimates in the downstream LGD-triangle becomes questionable, which compromises the tractability of the whole approach.

\subsubsection{The structural over-parametrisation of a ROT-model}
\label{app:ROT_demerits_overParams}

It is well-recognised in literature that a ROT is generally over-parametrised relative to the available data. Consider that a ROT-based model has a large number of parameters (or cells) even in a relatively small triangle, which means that the model is too complex for the amount of available data.
In particular, \citet{verrall1994} showed that allowing the development factors of a ROT to vary freely across rows and columns can lead to a ROT-parametrisation that is "far too large" in order to be statistically supportable. The practical implication hereof is that each observed cell exerts disproportionate influence on the estimated development factor, especially as the parameter count grows whilst the underlying dataset remains relatively small. In turn, this feature often prompts the need for further smoothing methods, which complicates the overall ROT-based implementation.

Considering an $n\times n$ square matrix of a ROT, the number of aggregated observations in the upper triangle alone is $n(n+1)/2$, before any forecasting can even occur. For example, in a $60\times 60$ triangle, the upper triangle consists of $ 1,830$ cells that must be estimated. Forecasting the lower triangle then requires a further $1,770$ estimates, resulting in a total of $3,600$ estimated values for a single triangle. We consider this total to be an exceptionally large number of parameters relative to the size of the available data, leading to demonstrable data sparsity in many of the triangle cells. This complexity is further compounded by segmentation, wherein separate ROT-models are typically estimated within specific sub-portfolios (or segments) of the broader portfolio towards risk-sensitising the eventual LGD-estimates. Doing so multiplies the effective parameter count and exacerbates the concerns around estimation error, parameter instability, and overall model risk.

\subsubsection{Recursive forecasting and error propagation in the ROT-framework}
\label{app:ROT_demerits_recursive}

In the ROT-framework, the recursive estimation of development factors and the subsequent forecasts in the lower triangle is prone to a form of error propagation that can occur in at least two ways. These development factors refer either to $\alpha_{v,m'}$ from \autoref{eq:devFactor_PW} or to $\beta_{v,m}$ from \autoref{eq:devFactor_LS} that use a rolling $k$-month cohort window. Firstly, the recursive structure implies that any estimation errors in early maturities propagate horizontally across subsequent maturities, since each forecasted cell is obtained by multiplying the previous value (itself either observed or forecasted) with an estimated development factor. Secondly, estimation errors can also propagate vertically across cohorts since development factors are iteratively re-estimated over a rolling window that includes previously forecasted values. Consequently, any bias introduced at an earlier stage is carried forward into all later stages of completing the ROT.

\begin{figure}[ht!]
    \centering
    \includegraphics[width=0.9\linewidth, height=0.5\textheight]{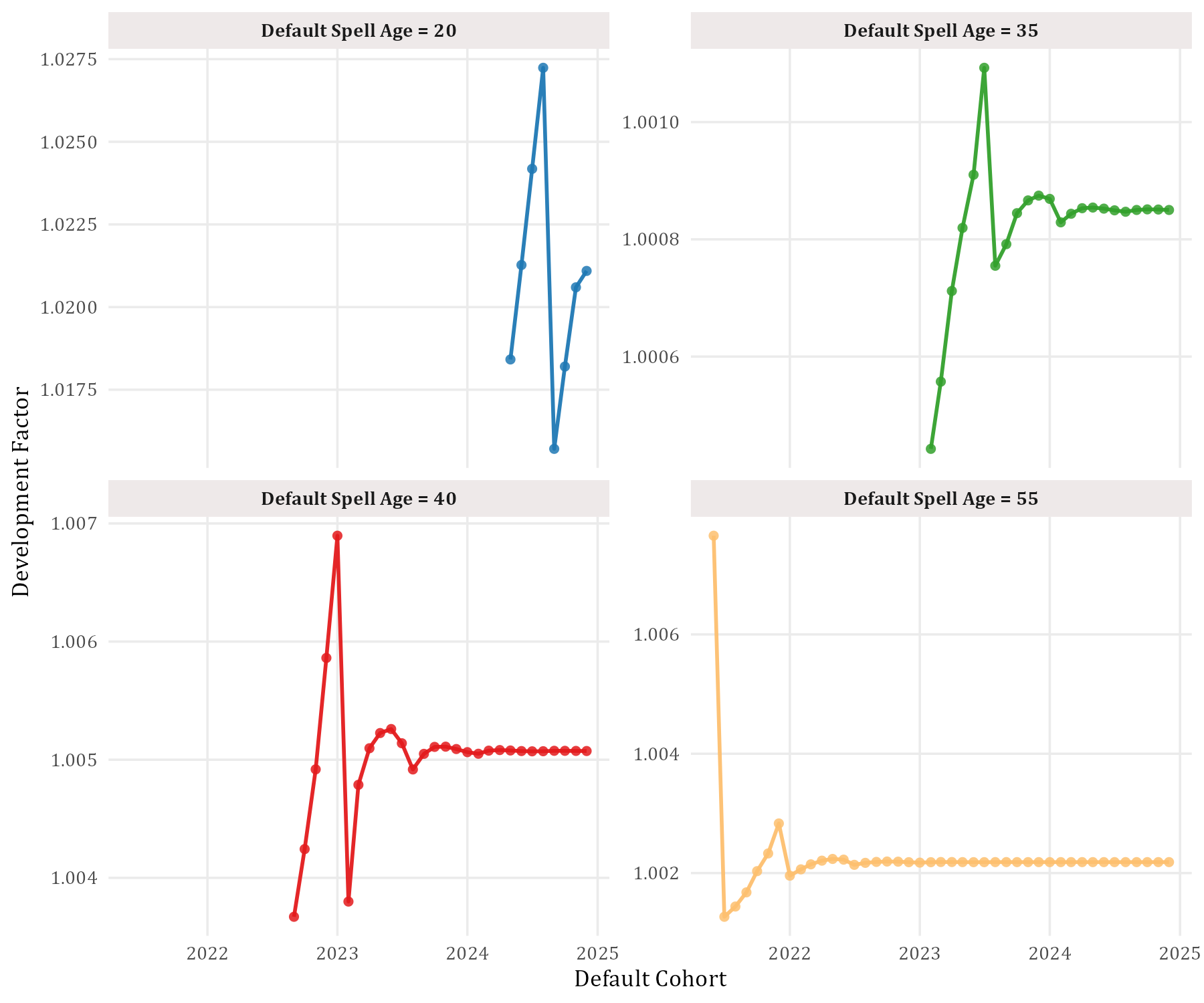}
    \caption{Estimated development factors $\alpha_{v,m'}$ from the PW-triangles, shown across default cohorts $v$ for selected default spell ages $m'$ across the panels.}
    \label{fig:ROT_DevFactors}
\end{figure}

Furthermore, the ROT-approach relies on the assumption that development factors are stable across default cohorts. This assumption allows for using the observed historical behaviour in the upper triangle to forecast the missing values in the lower triangle. Doing so implicitly assumes that write-off and loss dynamics are invariant over both time and across default cohorts. If this assumption is violated, perhaps due to changes in the credit policy, portfolio composition, or the prevailing macroeconomic conditions, then the development factors can become compromised. As an illustration, we examine the development factors $\alpha_{v,m'}$ for the PW-triangle over a span of default cohorts $v\in \{t_1,\dots,t_n\}$ for a few selected spell ages $m'\in\{ 20, 35, 40, 55\}$, as shown in \autoref{fig:ROT_DevFactors}. Fewer factors are available for smaller $m'$ values due to the triangular nature of a ROT. Evidently, sharp jumps occur in these factors and these jumps seem to occur inconsistently across cohorts before some of them start to converge. This phenomenon suggests that these factors are indeed unstable at those periods and are unlikely to be driven by a single systematic factor.

\subsubsection{Two-stage ROT-based LGD-values vs the mean empirical loss rates}
\label{app:ROT_demerits_divergence}

As explained in \autoref{sec:2stage_ROT_LGD}, the final set of ROT-based LGD-estimates over time $v$ is taken from the last column in the $G$-matrix per industry practice, i.e., $\mathrm{LGD}_v=G_{v,m}$ for $m=o$. This set constitutes a time series of supposedly "empirical" LGD-estimates, though we shall argue that this is not the case.
For any particular default cohort $v$, the $\mathrm{LGD}_v$ value from the ROT-model differs mathematically and structurally from simply taking the mean of the realised loss rates $l_i,i=1,\dots,n_v$, as calculated using \autoref{eq:realisedLossRate}, and aggregated as described in \autoref{sec:results}. By virtue of its construction, the $G$-matrix aggregates loss data through the PW-- and LS-triangles, which already assumes that the \textit{expected} LGD can be decomposed into two such triangles. Given the expected nature of this PW-LS decomposition, each resulting $G_{v,m}$ serves as a prediction in and of itself for any $v$ and $m$, and cannot be treated as a true empirically-observed loss from some underlying random variable.
Another point of consideration is that right-censored accounts are included (as write-offs) in the estimation of the entries in the LS-triangle, which further affects each downstream $G_{v,m}$ value. However, these right-censored cases lack an outcome by definition, which means that the associated loss cannot (yet) be observed empirically. Though certainly risk-prudent, simply assigning these right-censored cases as write-offs presumes future knowledge. This assumption injects bias into the eventual $G_{v,m}$ estimates, and can therefore compromise the final set of $\mathrm{LGD}_v$ estimates.

The ROT-model further assumes a fixed outcome period during which time maturities are observed, which means that observations beyond this horizon are excluded by definition. Varying this outcome period can alter each resulting $G_{v,m}$ estimate in the lower triangle, which highlights the dependence between the downstream $\mathrm{LGD}_v$ estimates (that are supposedly `empirical') and the ROT's parametrisation. Comparing $\mathrm{LGD}_v$ with the mean realised loss rate at the same $v$ represents two fundamentally different `lenses' through which the loss data can be viewed. However, the ROT-approach essentially amounts to using a model to produce "actual" LGD-outcomes in the upper triangle, which are then used as inputs into further forecasting LGD-values in the lower triangle. This circularity underscores the intractability of interpreting the ROT-derived $\mathrm{LGD}_v$ values as empirical observations. For these reasons, we posit that the latter of these two lenses (i.e., the mean realised loss rate) is the correct approach to observing true empirical loss rates.

\begin{figure}[ht!]
    \centering
    \includegraphics[width=0.8\linewidth, height=0.4\textheight]{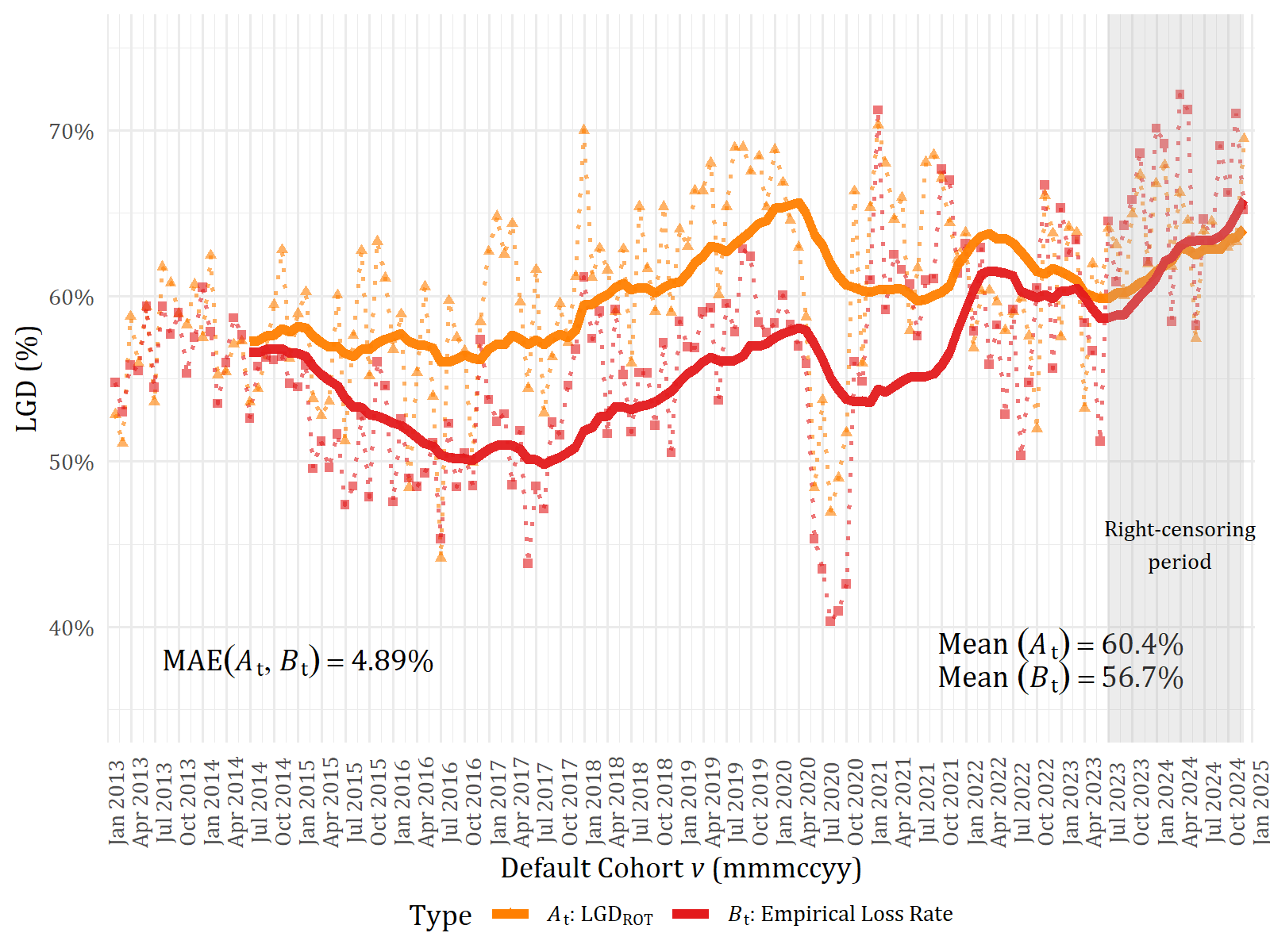}
    \caption{ROT-derived 'empirical' $\mathrm{LGD}_v$ values vs the mean of the realised loss rates $l_i$, calculated using \autoref{eq:realisedLossRate}. We impose an 18-month balance-weighted rolling average onto each time series to summarise the visual trends. The mean values of each time series are overlaid, as are the MAE-values between each series. The grey box indicates the period during which time right-censoring is material, thereby affecting the mean empirical loss rate.}
    \label{fig:ROT_LGDsOverTime}
\end{figure}

It is therefore unsurprising that a divergence exists between the $\mathrm{LGD}_v$ values and the true mean empirical loss rates over time $v$, as shown in \autoref{fig:ROT_LGDsOverTime}. To summarise the overall visual trend, we impose an 18-month rolling average onto each time series, which are shown in solid lines. Evidently, we observe both under-prediction and over-prediction when comparing $\mathrm{LGD}_v$ to the mean empirical rate, with the former attaining a mean value of 60.4\%  vs the grand mean empirical loss rate of 56.7\%. The ROT-based LGD values are therefore both mathematically and demonstrably different to the mean empirical rate.

\subsection{Selected variables within the components of a regression-based LGD-model}
\label{app:InputSpaces}

We follow a self-styled \textit{thematic} variable selection process using repeated regressions across themed subsets of input variables. This rather interactive process is guided by domain expertise, model parsimony, statistical significance, goodness-of-fit (GoF), multicollinearity, and model performance on the validation set $\mathcal{D}_\mathrm{V}$, as partially covered by \citet{heinze2018variable}. Candidate variables are first grouped into clearly defined themes, whereafter they are screened and assessed using a few measures across the aforementioned aspects. Specifically, variables are categorised into four themes: macroeconomic factors, delinquency dynamics, portfolio-level aggregates, and account-level behavioural characteristics. Within each theme, we formulate sub-themes in line with the scientific method that posits specific questions to be investigated, each of which is accompanied by a data-driven conclusion following some analysis. An example hereof is the identification of the optimal lag-order for a given macroeconomic variable, which forms the basis of a particular sub-theme and associated question: \textit{which lag-order is the `best'}?

Given a sub-theme, we regress the outcome variable onto each candidate variable on its own in forming a set of \textit{single-factor} models. These models, and by extension, the underlying input variables, may then be assessed using a few measures. Regarding model parsimony, \citet{Akaike1998} explained that a parsimonious model uses the smallest number of inputs relative to the sample size, whilst achieving the maximum GoF-value. This aspect may be measured for each single-factor model using the \textit{Akaike Information Criterion} (AIC): lower values indicate greater GoF. The results are compiled and the models are then ranked according to their AIC-values in a descending order, which implies a rank ordering of the corresponding input variables. The ranked AIC-values may be graphed and one may identify an "elbow point" beyond which changes in the AIC-values become vanishingly small. The ranked variables preceding this elbow point are retained for further analysis, and we discard variables beyond this point given their relatively negligible contribution to GoF.

\begin{figure}{ht!}
    \centering
    \includegraphics[width=0.7\linewidth, height=0.5\textheight]{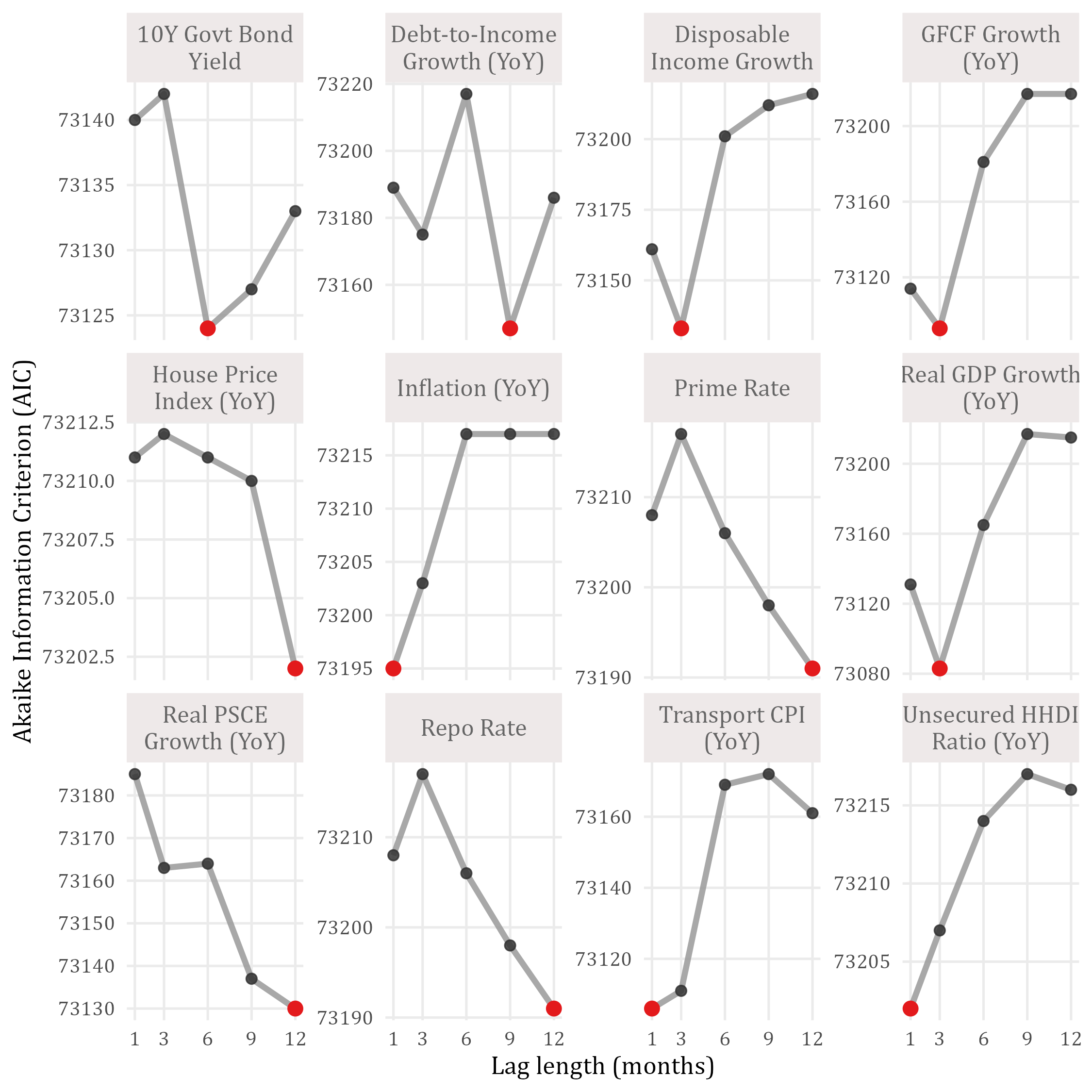}
    \caption{AIC by lag-order across single-factor models using macroeconomic variables. The abbreviated variables include the 10-year government (govt) bond yield, real personal consumption expenditure (PSCE) growth rate, real gross domestic product (GDP) growth rate, gross fixed capital formation (GFCF) growth rate, unsecured household debt-to-income (HHDI) growth rate, and the policy rate (Repo).}
    \label{fig:lagOrder_AIC}
\end{figure}

Input variables with time-dependent but systematic behaviour, such as macroeconomic variables and portfolio-level aggregates, are subjected to an additional lag-order selection procedure. This procedure identifies the most informative lag-order of a given variable, based on optimising an underlying criterion such as the AIC across single-factor models. Applying this procedure can help isolate delayed effects of such time-dependent systematic variables on the outcome variable. An example of such a delayed effect is the transmission time of changes in the central bank policy rate across the broader macroeconomic landscape, which may very well affect loan delinquency.
Within a given set of candidate lag-orders of a particular variable, each lag-order $k$ is evaluated independently by fitting a single-factor model using the $k$-lagged variant of that variable in predicting the outcome. The AIC-values of the resulting set of models are then graphed, as shown in \autoref{fig:lagOrder_AIC} for various macroeconomic variables, and the optimal lag-order is chosen by selecting the best (lowest) AIC-value.

At the end of each theme, we assess the magnitude of multicollinearity amongst the selected input variables, and deselect affected variables accordingly. As discussed by \citet{belsley1980}, multicollinearity is an effect whereby two or more inputs exhibit substantial linear dependence within a GLM. Testing the degree of multicollinearity is crucial since the sampling variance of the coefficient estimates within a GLM can become inflated amidst collinear variables, which may lead to unstable inference. A common approach to gauging multicollinearity is to calculate the \textit{variance inflation factor} (VIF) of each input and compare it across inputs. This factor, denoted as $VIF_j$ for each input $j$, is defined as $(1-R^2_j)^{-1}$, where $R^2_j$ is the coefficient of determination obtained by regressing the input $x_j$ on all other inputs. As $R^2_j \rightarrow1$, the variance of the associated coefficient estimator $\hat{\beta}_j$ increases, which indicates stronger multicollinearity. A VIF-value of one indicates no multicollinearity, whereas larger VIF-values reflect increasing collinearity and reduced precision of coefficient estimates. That said, no universally accepted VIF-threshold exists beyond which multicollinearity becomes problematic. \citet{de2015impact} studied the effects of varying the VIF-threshold when methodically deselecting variables in an application credit scorecard model, itself based on logistic regression. They found that imposing overly strict VIF-thresholds (e.g., $\leq2.5$ or $\leq5$) can unnecessarily erode discriminatory power, and becomes a questionable practice when working with larger datasets. Accordingly, and given the relatively large size of our own dataset, we adopt a VIF-threshold of 10 based on industry experience, and deselect variables accordingly. In so doing, we deselect one variable in the PWGD-model, whilst deselecting two variables in the LGW-model.

Thereafter, we apply a screening procedure onto the selected input variables, which further curtails the input space towards achieving true model parsimony. This screening procedure is based on the mean Shapley-value per variable, as approximated by SHAP-values; see \citet{strumbelj2014}. Each model-agnostic SHAP-value quantifies the importance of the associated variable by assessing its marginal contribution to the model output, relative to a common baseline prediction. The variables are then ranked in a descending order given these mean SHAP-values, whereafter one may again identify an "elbow point" beyond which the mean SHAP-values do not change by much. By implication, the underlying variables with mean SHAP-values beyond this point are discarded on the basis of contributing little to the predictions of the model. Accordingly, we select the top 15 variables.
Finally, we apply expert judgment in curating the final list of inputs and retain strategically-important variables such as the debt counselling flag \texttt{[DC\_Ind]}. This final list of variables is compiled in \autoref{tab:featuresDescription} across the two component models (PWGD \& LGW).

\begin{longtable}{p{3.7cm} p{9.3cm} p{2cm}}
\caption{The selected input variables mapped across the two component models (PWGD \& LGW) within the overarching LGD-model. Subscripts $[\mathrm{a}]$ denote loan account-level variables, $[\mathrm{p}]$ are portfolio-level inputs, and $[\mathrm{m}]$ represent macroeconomic covariates. A colon (\texttt{:}) between variable names denotes an interaction term, while an asterisk (\texttt{*}) indicates a variable whose measurement window differs across the two component models.}
\label{tab:featuresDescription} \\
\toprule
\textbf{Variable} & \textbf{Description} & \textbf{Models} \\ 
\midrule
\endfirsthead
\caption[]{(continued)} \\
\toprule
\textbf{Variable} & \textbf{Description} & \textbf{Models} \\ 
\midrule
\endhead
\midrule \multicolumn{3}{r}{\textit{Continued on next page}} \\
\endfoot
\bottomrule
\endlastfoot

\footnotesize{\texttt{Age\_WoE}$_{[\mathrm{a}]}$} & \footnotesize{Age of the loan, measured as the number of months elapsed between the origination date and the observation date. This variable is transformed using the \textit{weights-of-evidence} (WoE) technique.} & \footnotesize{PWGD} \\

\footnotesize{\texttt{Arrears\_Bin}$_{[\mathrm{a}]}$} & \footnotesize{Ordinal indicator representing the arrears bin at month end, defined as 0,1,2,3, or 4+ arrears.} & \footnotesize{PWGD} \\

\footnotesize{\path{D7_Ind:IncomeToLimit_Wins}$_{[\mathrm{a}]}$} & \footnotesize{The interaction effect between the binary indicator variable that flags the cases wherein Directive 7/2015 of the SARB holds relating to distressed restructures, and the winsorised annual income-to-limit ratio.} & \footnotesize{PWGD} \\

\footnotesize{\texttt{DC\_Ind}$_{[\mathrm{a}]}$} & \footnotesize{Binary indicator denoting whether the account is in debt counselling.} & \footnotesize{PWGD} \\

\footnotesize{\texttt{DefSpell\_Age}$_{[\mathrm{a}]}$} & \footnotesize{Total length (in months) of the account's default spell, measured from the start of default until resolution. This variable is represented as discretised risk bins (\texttt{DefSpell\_Age\_Bin}) in the LGW model and transformed using WoE-encoding (\texttt{DefSpell\_Age\_WoE}) in the PWGD model.} &
\footnotesize{LGW, PWGD} \\

\footnotesize{\texttt{DefSpell\_Num}$_{[\mathrm{a}]}$} & \footnotesize{Ordinal indicator representing the borrower's current default episode, equivalent to the cumulative number of transitions into a default state, observed up to the reporting date.} & \footnotesize{PWGD} \\

\footnotesize{\texttt{Delinq\_g0\_Prop\_12\_WoE}$_{[\mathrm{p}]}$} & \footnotesize{Non-defaulted fraction of the portfolio with any degree of delinquency beyond g$_{\mathrm{0}}$ = 0, lagged by 12 months and WoE-encoded.} & \footnotesize{PWGD} \\

\footnotesize{\path{Delinq_g0_Prop_12:DC_Ind}$_{[\mathrm{a}]}$} & \footnotesize{The interaction effect between the binary indicator variable that flags whether the account is in debt counselling, and the non-defaulted fraction of the portfolio with any degree of delinquency beyond zero payments in arrears, lagged by 12 months.} & \footnotesize{LGW} \\

\footnotesize{\texttt{IncomeToLimit\_Wins\_WoE}$_{[\mathrm{a}]}$} & \footnotesize{Ratio of the borrower's annual income to the approved credit limit. This variable is winsorised and subsequently WoE-encoded. } & \footnotesize{PWGD} \\

\footnotesize{\texttt{Legal\_Ind}$_{[\mathrm{a}]}$} & \footnotesize{Binary indicator denoting whether the account is subject to legal recovery proceedings.} & \footnotesize{PWGD} \\

\footnotesize{\texttt{M\_Disp\_Inc\_Growth\_*}$_{[\mathrm{m}]}$} &
\footnotesize{Year-on-year growth rate in household disposable income. The quarterly series is interpolated to a monthly frequency, lagged by 3 months in the PWGD-model and 12 months in the LGW-model.} &
\footnotesize{LGW, PWGD} \\

\footnotesize{\texttt{M\_DTI\_Growth\_YoY\_9\_WoE}$_{[\mathrm{m}]}$}&
\footnotesize{Year-on-year growth rate in the household debt-to-income ratio. The quarterly series is interpolated to a monthly frequency, lagged by nine months, and transformed using WoE-encoding.}&\footnotesize{PWGD}\\

\footnotesize{\texttt{M\_GDP\_Growth\_YoY\_3\_WoE}$_{[\mathrm{m}]}$} &
\footnotesize{Year-on-year growth rate in real gross domestic product (GDP). The quarterly series is interpolated to a monthly frequency, lagged by three months, and transformed using WoE-encoding.} & \footnotesize{PWGD} \\

\footnotesize{\texttt{MB\_Ind}$_{[\mathrm{a}]}$} &
\footnotesize{Binary indicator denoting whether the borrower is main banked (MB) with the Bank or non-main banked (NMB).} &
\footnotesize{LGW, PWGD} \\

\footnotesize{\texttt{NoPayment\_Count\_4}$_{[\mathrm{a}]}$} & \footnotesize{The rolling count of missed payments over the previous four months.}&
\footnotesize{PWGD} \\

\footnotesize{\texttt{Num\_Consec\_No\_PMT\_4}$_{[\mathrm{a}]}$} & \footnotesize{The rolling count of consecutive missed payments over the previous four months.}&
\footnotesize{PWGD} \\

\footnotesize{\path{Num_Consec_No_Pmt*:DefSpell_Age_Bin}$_{[\mathrm{a}]}$} &
\footnotesize{An interaction term between the number of consecutive non-payments and \texttt{DefSpell\_Age\_Bin}. The number of consecutive non-payments is measured over a 4-month window in the PWGD model and a 9-month window in the LGW model.} &
\footnotesize{LGW, PWGD} \\

\footnotesize{\path{Num_Consec_No_Pmt_9:Util_Wins_Prev}$_{[\mathrm{a}]}$} & \footnotesize{An interaction effect between the number of consecutive missed payments, measured over a 9-month window, and the previous month's winsorised utilisation ratio.} & \footnotesize{LGW} \\

\footnotesize{\texttt{Redefault\_Ind}$_{[\mathrm{a}]}$}&
\footnotesize{Binary indicator denoting whether the account has redefaulted; i.e. 1 if the account has experienced more than one default episode, and 0 otherwise.} & \footnotesize{LGW} \\

\footnotesize{\texttt{Util\_Dir\_6}$_{[\mathrm{a}]}$} & \footnotesize{Categorical variable describing the short-term directional behaviour of the utilisation over a six-month rolling window, with the categories being 1) Increasing, 2) Decreasing, 3) Milling and 4) Missing.} & \footnotesize{PWGD} \\

\footnotesize{\texttt{Util\_Ratio\_Wins\_12}$_{[\mathrm{a}]}$} &
\footnotesize{Winsorised ratio of average utilisation over the most recent six months to that of the preceding six months. Values greater than one indicates increasing utilisation, values below one indicates decreasing utilisation, and values near one indicate stable utilisation.} &
\footnotesize{LGW} \\

\footnotesize{\texttt{Util\_Wins\_Prev}$_{[\mathrm{a}]}$} &
\footnotesize{Utilisation, defined as the ratio of \texttt{Balance\_Wins} to \texttt{Limit\_Wins}. This variable is winsorised and lagged by one month.} &
\footnotesize{LGW, PWGD} \\

\end{longtable}
\subsection{Two diagnostics on the components of a regression-based LGD-model}
\label{app:Diagnostics_2Stage_Regression}

As detailed in \autoref{sec:2stage_components}, the PWGD models are evaluated using three diagnostics. First, the write-off resolution rate assesses the accuracy of the PWGD-models across all default cohorts in the sampling window. Second, ROC-analyses are used to quantify the overall discriminatory power of the models. Lastly, a Brier score (BS) is estimated to quantify the overall accuracy of the model and is analogous to a mean squared error. We now extend the last two diagnostics to a default cohort-level, which enables the assessment of discriminatory power and the accuracy of the model over time. A well-calibrated model should have consistently high discriminatory power and accuracy across most default cohorts, thereby affirming its ability to predict write-off amidst various changes in the macroeconomic background and in the portfolio itself.

\begin{figure}[ht!]
    \centering
    \includegraphics[width=0.75\linewidth,height=0.44\textheight]{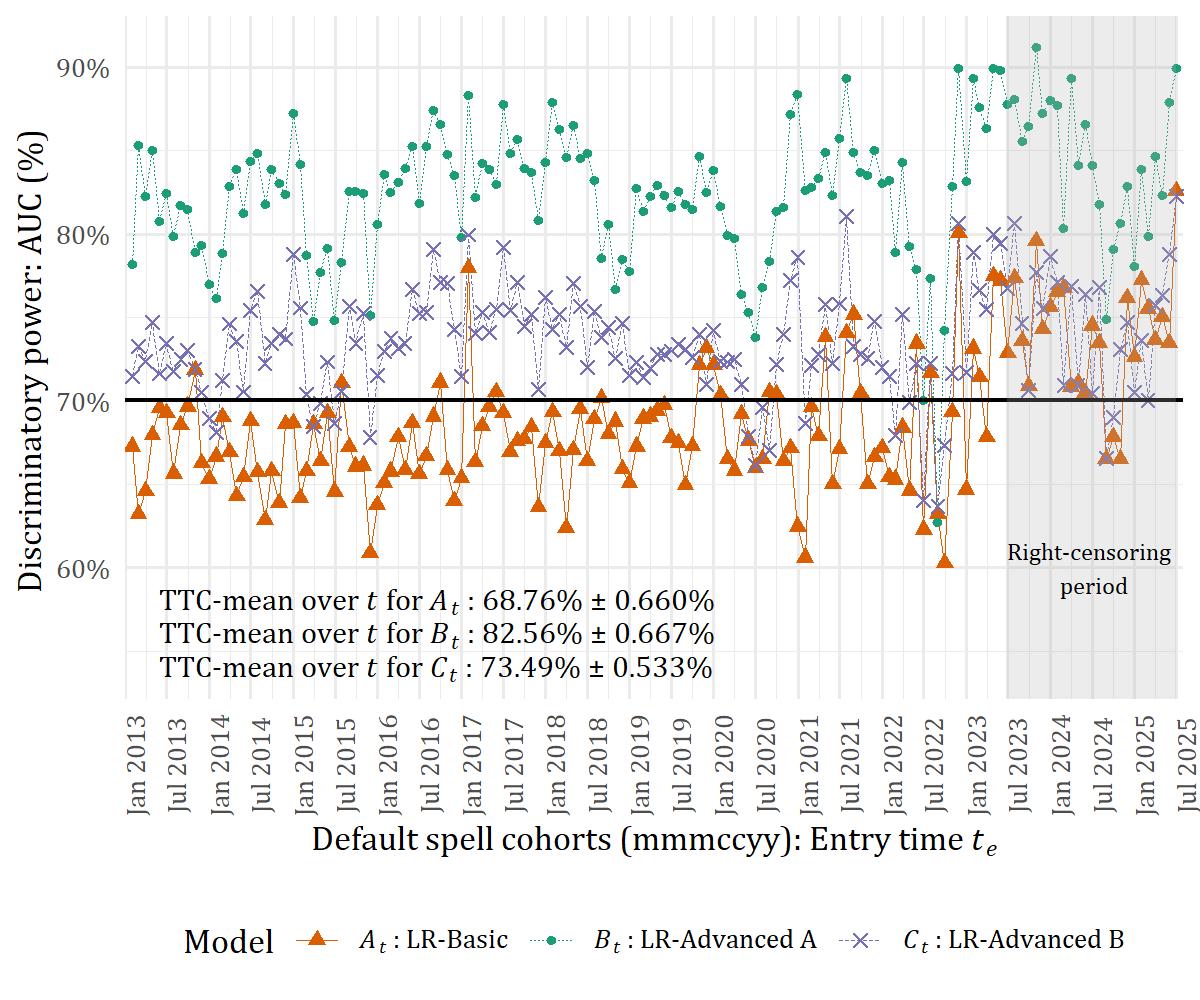}
    \caption{The cohort-level AUC-values estimated across default cohorts $v$ for the basic, and advanced Type A and B models, having used the full set $\mathcal{D}$. The \textit{through-the-cycle} (TTC) mean AUC is shown for each model, accompanied by a 95\% confidence interval that is calculated using z-scores.}
    \label{fig:AUC_CohortLevel}
\end{figure}

Conducting ROC analyses at the cohort-level simply requires stratifying a particular dataset $\mathcal{D}$ by each default cohort $v$, i.e., $\mathcal{D} = \bigcup_v \mathcal{D}(v)$. Thereafter, an ROC-analysis can be performed within each stratum. Cohort-level ROC-analyses are conducted for each PWGD-model 
across values of $v$, and the resulting cohort-level AUC-values are reported in \autoref{fig:AUC_CohortLevel} over $v$. The advanced type A model achieves the highest discriminatory power over most cohorts, followed by the advanced type B and, lastly, the basic model. These results attest of the more predictive input space of the advanced type A and B models, which engenders an improved ability to risk-rank predictions across broader systematic (or macroeconomic) effects.
Less credence is given to those AUC-values within the most recent periods during which right-censoring becomes increasingly prevalent, largely due to dwindling sample sizes in those datasets $\mathcal{D}(v)$ for $v\geq$ Jun-2023.

\begin{figure}[ht!]
    \centering
    \includegraphics[width=0.75\linewidth,height=0.44\textheight]{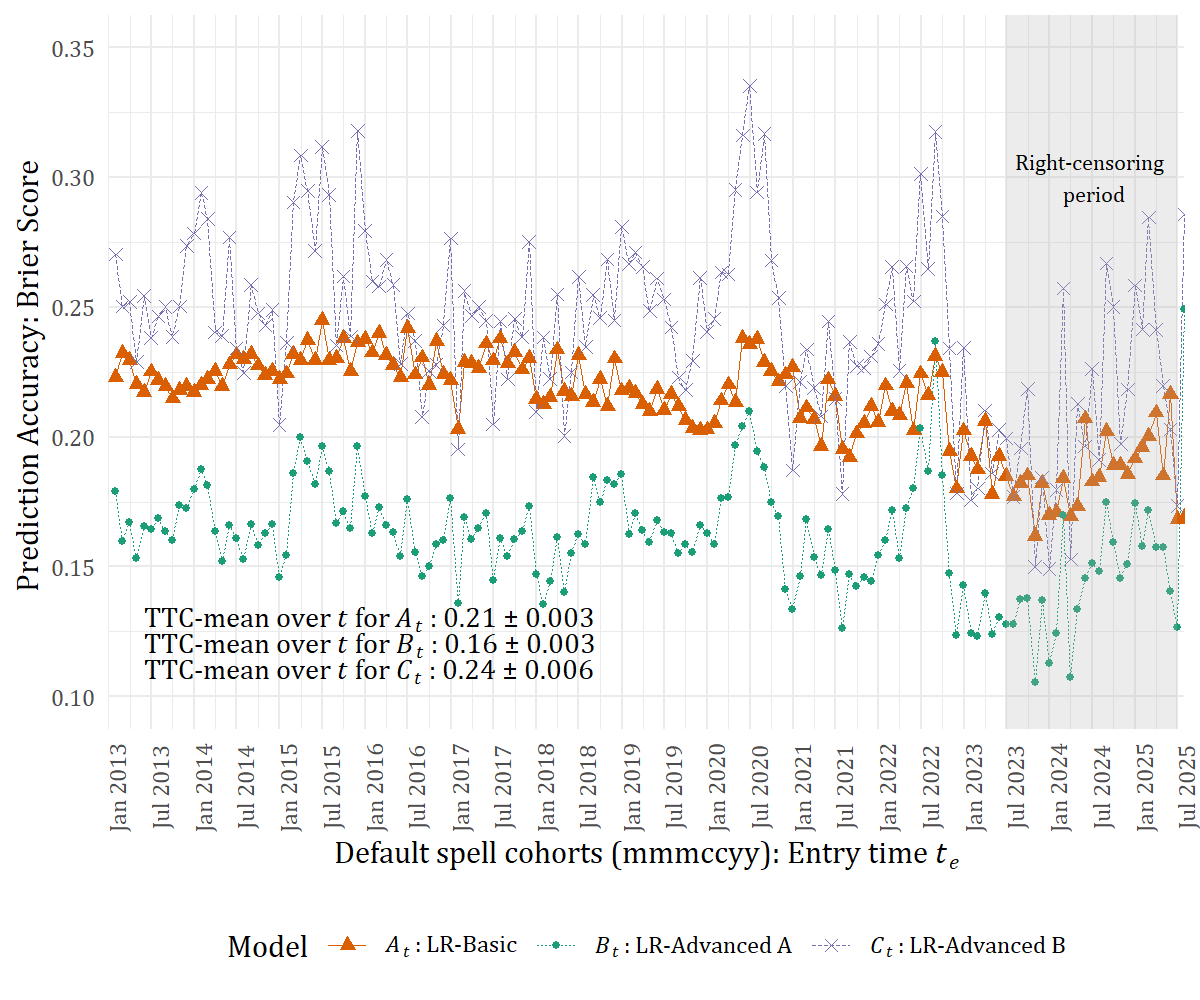}
    \caption{Cohort-level Brier Scores (BS) estimated across default cohorts $v$ for the basic, and Advanced Type A and B models, having using the full dataset $\mathcal{D}$. The TTC-mean BS is shown for each model and is accompanied by a 95\% confidence interval that is calculated using z-scores.}
    \label{fig:BrierScore_CohortLevel}
\end{figure}

Similarly to the cohort-level ROC-analyses, we extend the overall BS-statistic to a cohort-level by stratifying the full dataset $\mathcal{D}$ by default cohort $v$ and estimating the BS using data from each $\mathcal{D}(v)$. Again, we perform cohort-level BS-analyses for each PWGD-model across values of $v$, and show the resulting time series in \autoref{fig:BrierScore_CohortLevel}. The advanced Type A model is best able to predict write-off across most default cohorts, whilst the basic model also has commendable accuracy. The dichotomisation of the probability predictions within the advanced Type B model results in information loss, and therefore stymies the model's prediction accuracy across most of the default cohorts.

Both the cohort-level ROC and BS analyses reflect the trends in the overall ROC and BS results of the models, as reported in \autoref{sec:2stage_Regression}. The richer input spaces of the advanced Type A and B models ensure that the models have high discriminatory power across various default cohorts. The advanced Type A model also has the highest accuracy across all default cohorts, whilst the accuracy of the advanced type B model is adversely affected by its binary-valued predictions.

\subsection{A short procedure for dichotomising the PWGD-model using the Generalised Youden Index}
\label{app:PWGD_Dichotomisation}

The estimated write-off probability $\pi_i$ of default spell $i$ with inputs $\boldsymbol{x}_i$ is dichotomised into a binary value using an optimally-chosen cut-off $c^*\in[0,1]$, which is operationalised as $\mathbb{I}(\pi_i>c^*)$ using the indicator function $\mathbb{I}(\cdot)$. The selection of this threshold $c^*$ amongst all possible thresholds $c\in[0,1]$ is based on a certain trade-off between two types of misclassification. In formalising this trade-off, consider the notions of \textit{sensitivity} and \textit{specificity} that originate from ROC-analysis, as discussed by \citet{fawcett2006introduction}. In our context, sensitivity represents the probability of correctly identifying a write-off event (or a \textit{true positive}) and is given by 
\begin{equation} \label{eq:sensitivity}
    q(c)=\mathbb{P} (\pi_i>c \mid y_i = 1) \, ,
\end{equation}
where $y_i$ is the observed write-off outcome. Similarly, specificity represents the probability of correctly identifying a cured event (or a \textit{true negative}) and is defined as
\begin{equation} \label{eq.}
    p(c) = \mathbb{P}(\pi_i \leq c \mid y_i=0) \, .
\end{equation}

Both sensitivity and specificity clearly vary with $c$ and they capture the fundamental trade-off in binary classification: lowering $c$ generally increases sensitivity $q(c)$ while decreasing specificity $p(c)$, and vice versa. Furthermore, this trade-off becomes crucial since the underlying misclassification costs are asymmetric in the LGD-modelling context. Misclassifying a write-off as a cure (or a \textit{false negative}) leads to underestimation of losses, whereas incorrectly flagging a cure as a write-off (or a \textit{false positive}) results in greater loss estimates. In risk management, we would prefer false positives over false negatives in that we would rather overestimate losses than underestimate it in our predictions thereof.

Let $a>0$ denote the relative cost of a false negative compared to a false positive. Let $\phi$ represent the prevalence of write-off events in a sample, as calculated using the arithmetic mean across write-off outcomes $y_i,i\in\{1,\dots,n\}$. Both $a$ and $\phi$ are incorporated into a cost-sensitive performance criterion called the \textit{Generalised Youden Index} (GYI), which evaluates classification quality for a given threshold $c$ under asymmetric misclassification costs. Following the work of \citet{geisser1998comparing}, \citet{schisterman2008youden}, and \citet{kaivanto2008maximization}, the GYI is defined as
\begin{equation} \label{eq:GYI}
    J_a(c) = q(c) + \frac{1-\phi}{a\phi}p(c) - 1 \, ,
\end{equation}
with $J_a(c)\in [-1,1]$. The optimal cut-off $c^*$ is then given by
\begin{equation} \label{eq:GYI_optimal}
    c^* = \arg \max_c J_a(c).
\end{equation}

The parameter $a$ directly governs the weight assigned to the specificity $p(c)$ in the GYI. As $a$ increases, the weight assigned to $p(c)$ decreases. Hence, the optimal threshold $c^*$ decreases and the classifier becomes more liberal in predicting the positive class (write-off). In the limit of $a\rightarrow\infty$, we have $J_a(c)\rightarrow q(c)-1$, and the classifier almost always predicts the positive class as $c^*\rightarrow 0$. Furthermore, the GYI resolves to the classical Youden index when setting $a=(1-\phi)/\phi$, wherein $q(c)$ and $p(c)$ contribute equally to the index as a special case of the GYI.

\begin{figure}[ht!]
    \centering
    \includegraphics[width=0.7\linewidth,height=0.32\textheight]{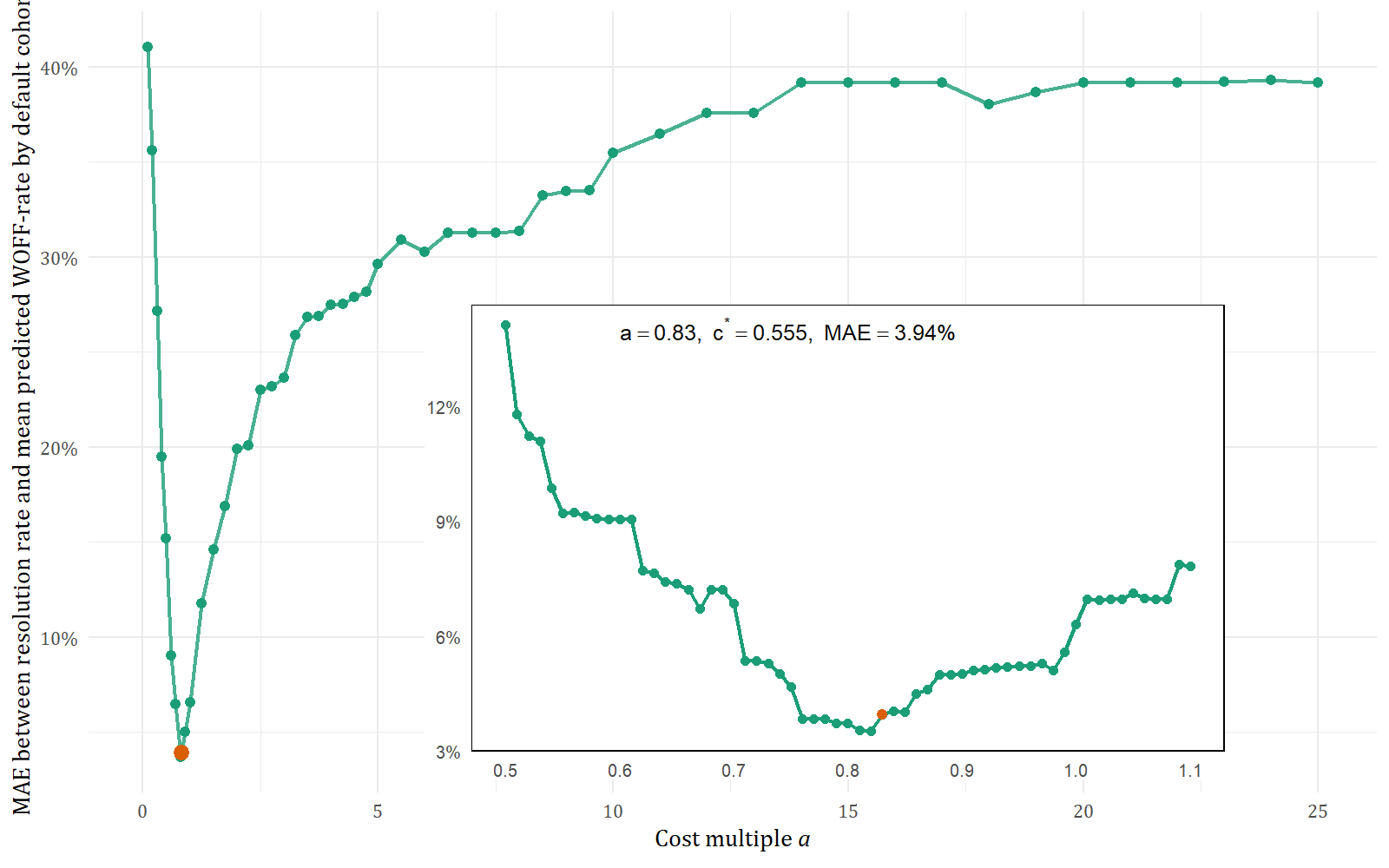}
    \caption{MAE-values calculated between the actual and the expected resolution rate, itself obtained from various dichotomisations of the PWGD-model, as embedded through a range of cost multiples $a$. The orange point highlights the select $a$ value, and associated metrics are annotated.}
    \label{fig:Opt_CutOff_MAE}
\end{figure}

\begin{figure}[ht!]
    \centering
    \includegraphics[width=0.7\linewidth,height=0.32\textheight]{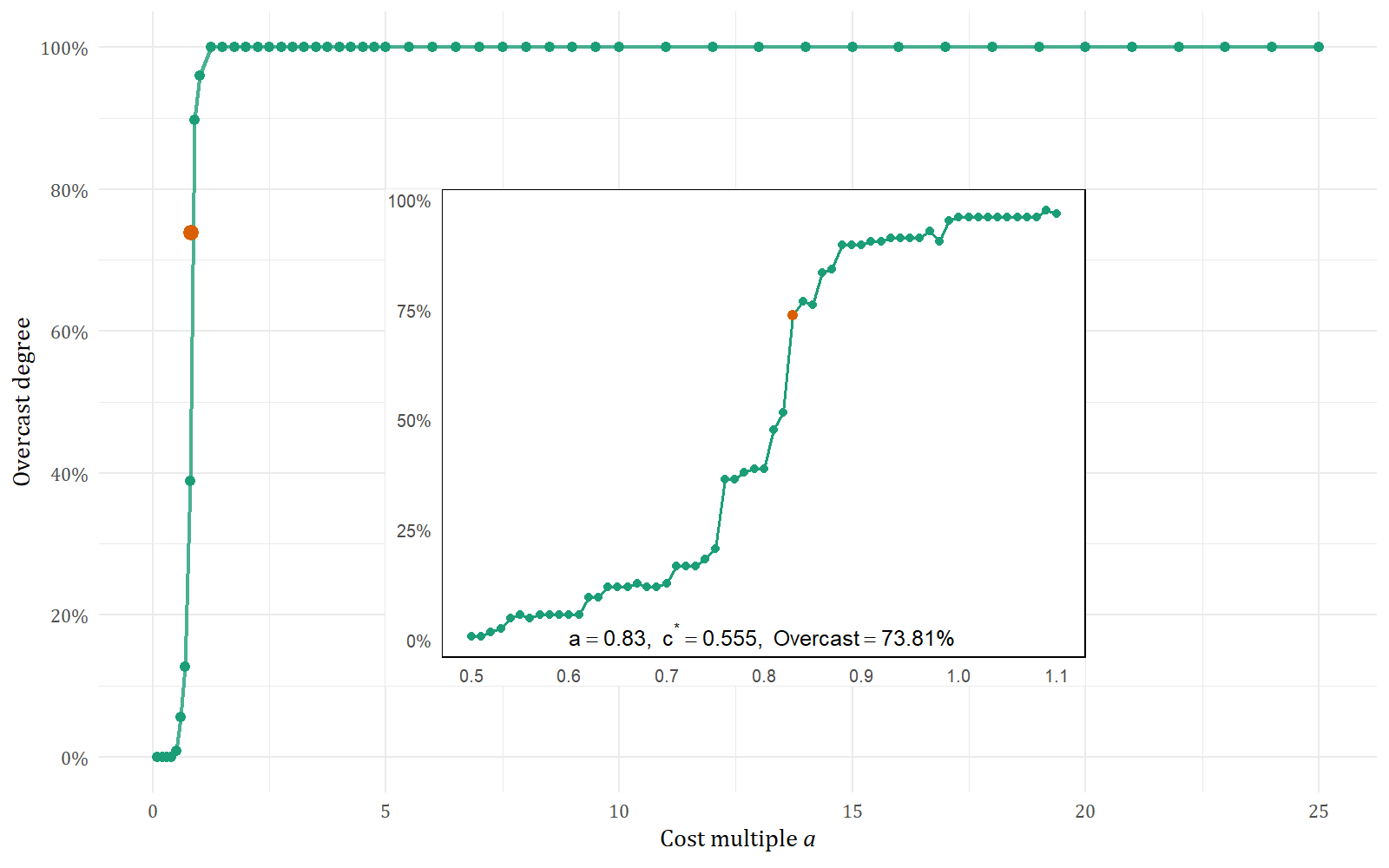}
    \caption{Overcast degrees calculated between the actual and the expected resolution rate, itself obtained from various dichotomisations of the PWGD-model, as embedded through a range of cost multiples $a$. Graph design follows that of \autoref{fig:Opt_CutOff_MAE}.}
    \label{fig:Opt_CutOff_OvercastDegreees}
\end{figure}

In finding $a$, we compare the actual write-off resolution rate with the expected rate obtained from the advanced PWGD-model, whose predictions are dichotomised using $c^*$ for a specific $a$. This comparison is repeated across various values of $a$, whereafter the associated MAEs and overcast degrees between the actual and expected write-off resolution rates are calculated. These measures (the MAE and overcast degree) are respectively shown in \autoref{fig:Opt_CutOff_MAE} and \autoref{fig:Opt_CutOff_OvercastDegreees}.
Although a cost multiple of $a=0.82$ yields a minimum MAE of 3.52\%, the associated overcast degree is relatively low at only 51.59\%. We would prefer a higher overcast degree in attaining more conservatism in the predictions, i.e., preferring over-prediction over under-prediction. Instead, a slightly higher value of $a=0.83$ is selected such that high prediction accuracy is balanced against increased conservatism. The associated MAE is thus 3.94\% and the associated overcast degree is 73.81\%. Accordingly, the chosen optimal cut-off is $c^*=0.555$.

\subsection{An analysis of expected vs empirical LGD-distributions}
\label{app:LGD_Distributions}

\begin{figure}[ht!]
    \centering
    \includegraphics[width=0.7\linewidth,height=0.41\textheight]{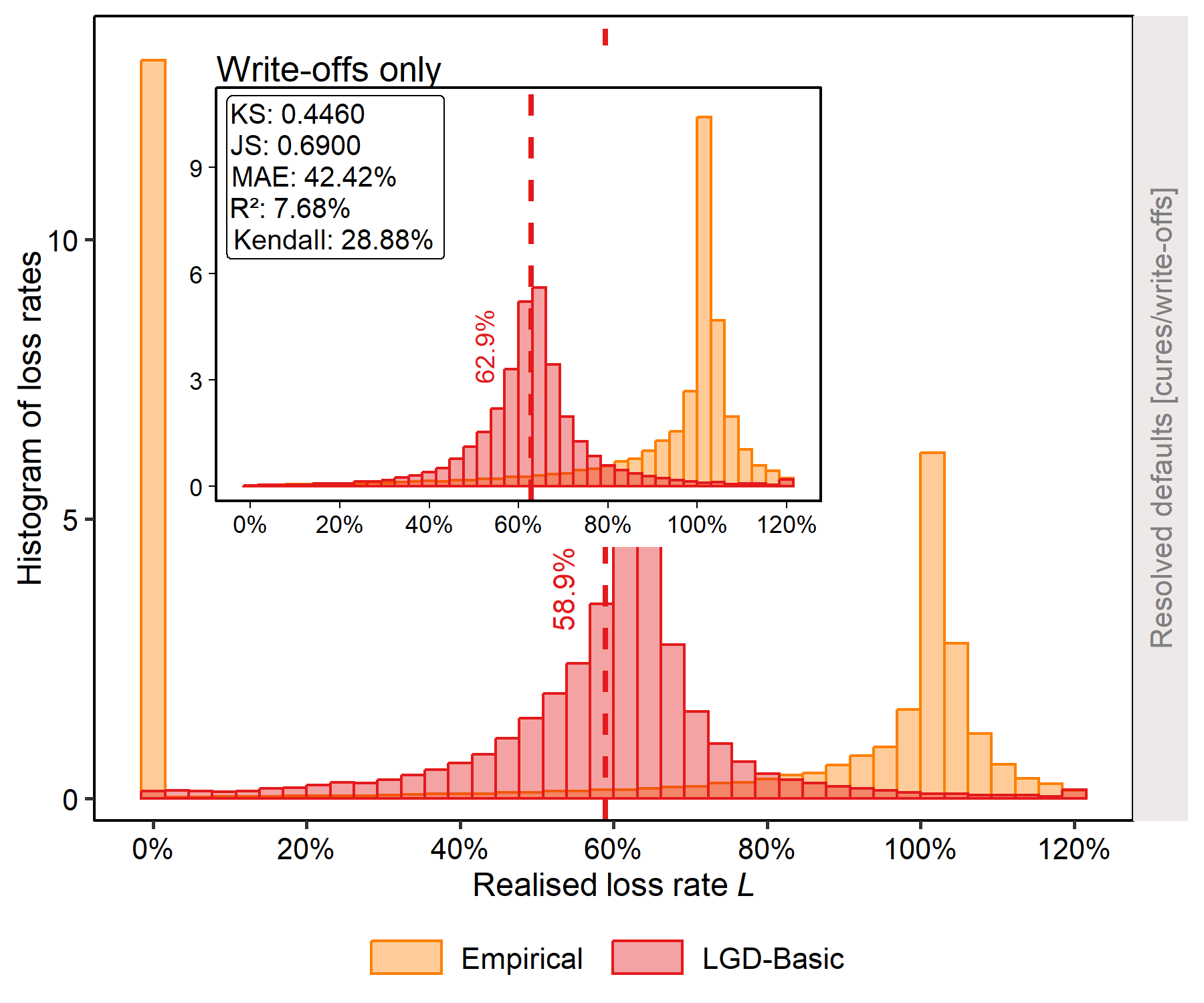}
    \caption{Empirical and expected LGD-distributions for resolved default spells. The expected LGD-values are reconstructed using the basic PWGD and basic LGW models, as scored on the full dataset $\mathcal{D}$. Each dashed line represents a mean predicted LGD-value, and annotation boxes summarise the distributional fit statistics.}
    \label{fig:EmpExp_LGD_basic}
\end{figure}

\begin{figure}[ht!]
    \centering
    \includegraphics[width=0.7\linewidth,height=0.41\textheight]{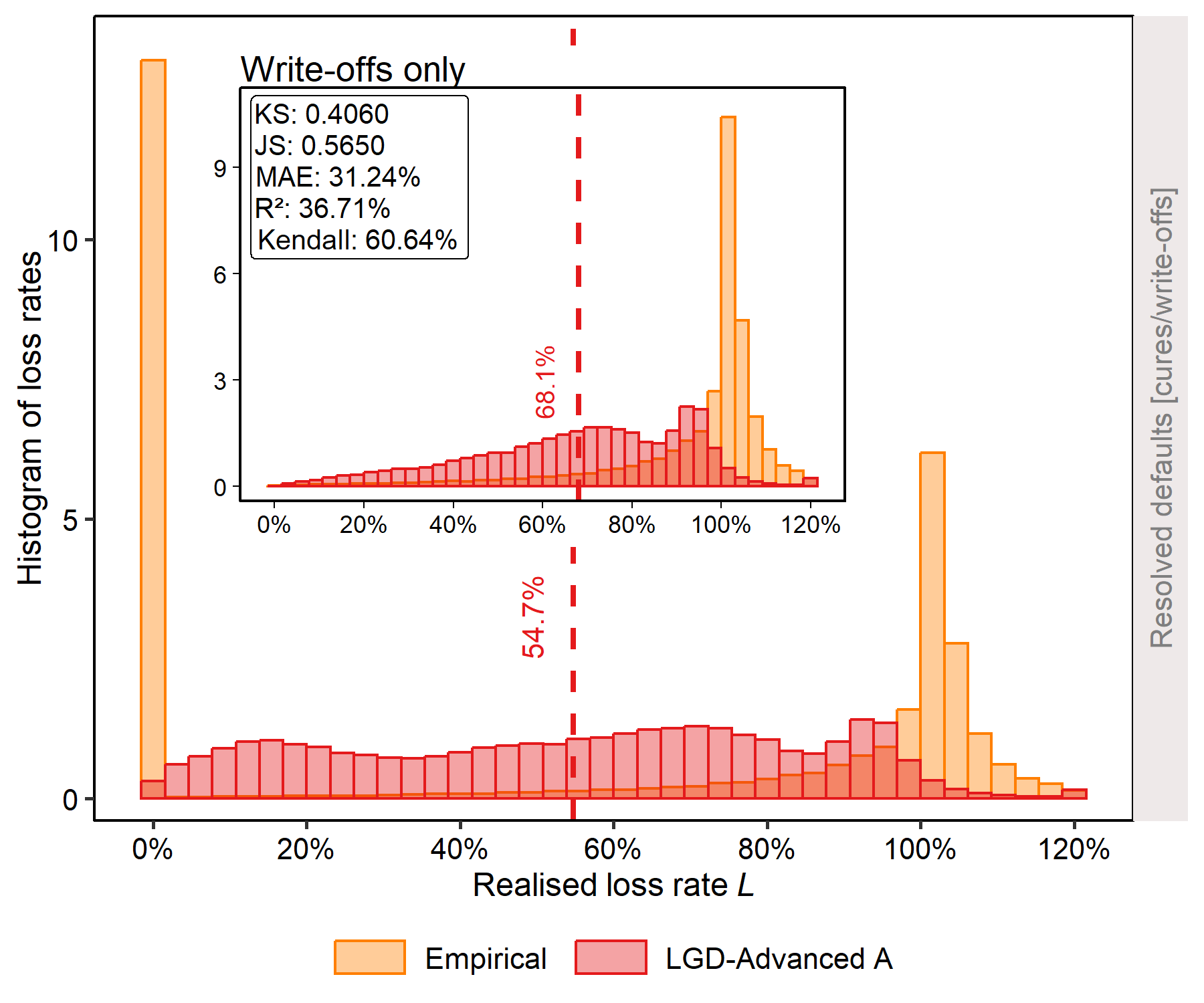}
    \caption{Empirical and expected LGD-distributions for resolved default spells. The expected LGD-values are reconstructed using the advanced Type A PWGD and advanced LGW models. Graph design follows that of \autoref{fig:EmpExp_LGD_basic}. }
    \label{fig:EmpExp_LGD_adv_TypeA}
\end{figure}

\begin{figure}[ht!]
    \centering
    \includegraphics[width=0.7\linewidth,height=0.41\textheight]{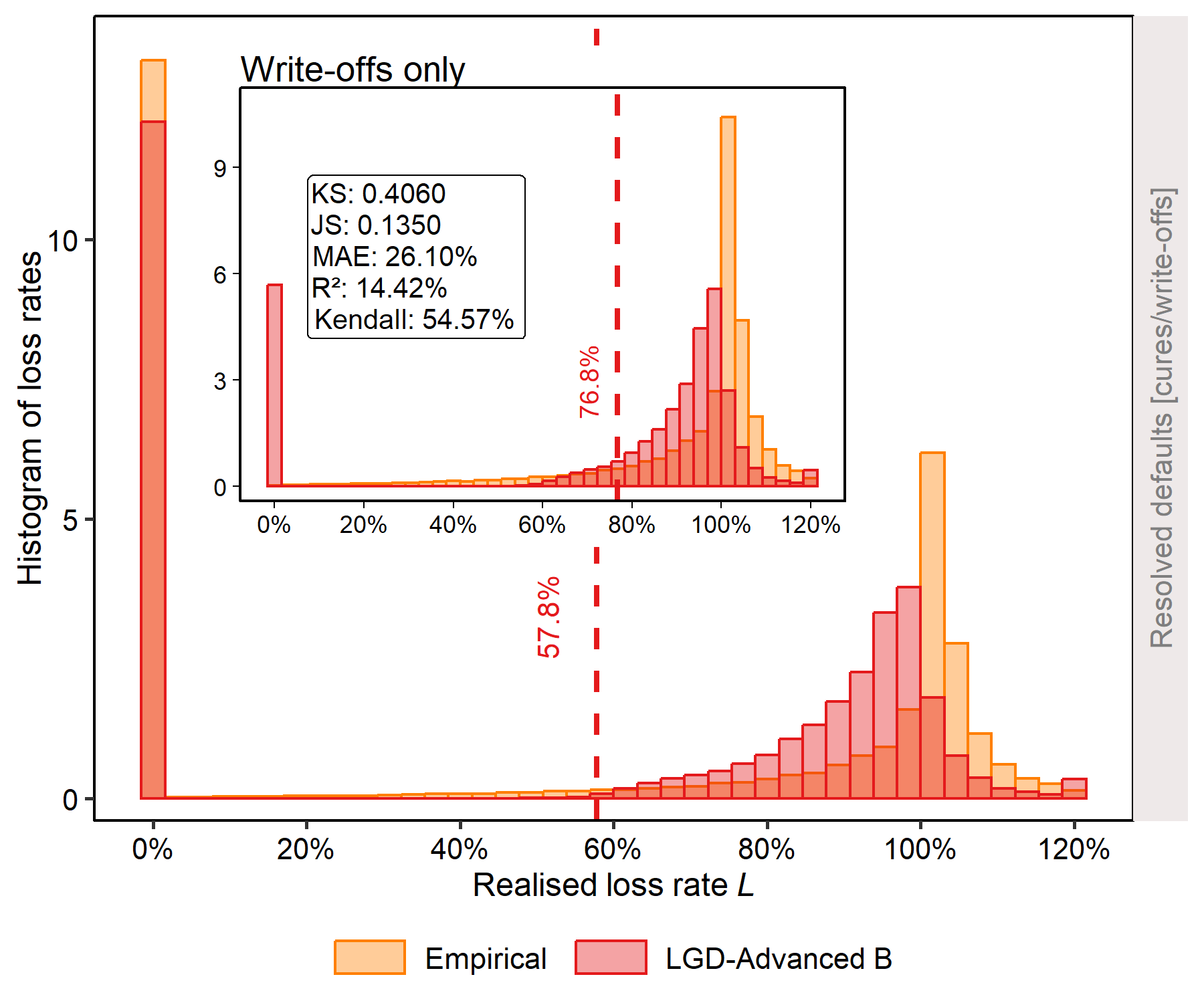}
    \caption{Empirical and expected LGD-distributions for resolved default spells. The expected LGD-values are reconstructed using the advanced Type B PWGD and advanced LGW models. Graph design follows that of \autoref{fig:EmpExp_LGD_basic}. }
    \label{fig:EmpExp_LGD_adv_TypeB}
\end{figure}

Consider the three LGD-models presented in the main text: 1) basic; 2) advanced Type A; and 3) advanced Type B. These models produce account-level LGD-estimates whose distributions can be graphed, as shown in \crefrange{fig:EmpExp_LGD_basic}{fig:EmpExp_LGD_adv_TypeB}, and compared against that of the empirical loss rates of all resolved default spells. Such a comparative distributional analysis provides insight not only into the predictive accuracy of an LGD-model, but also on its ability to reproduce the overall characteristics of the observed LGD-distribution.
Consider the basic LGD-model, which demonstrably fails to recover the pronounced right-skewness of the empirical LGD-distribution, as well as being unable to reproduce the large probability mass at zero. This deficiency is reflected in the distributional fit statistics, as annotated on the graphs, relative to the distributions of the other LGD-models. In particular, the \textit{Jensen-Shannon} (JS) divergence measure gives a relatively high value of 0.69 for the basic LGD-model, when compared to that of the other LGD-models: 0.565 for the advanced Type A model, and a remarkably lower value of 0.135 for the advanced Type B model. Lower JS-values are deemed to be better. Furthermore, the \textit{Kolmogorov-Smirnov} (KS) test statistics reflect a similar pattern in their values across the competing LGD-models, and rank-orders the LGD-models exactly the same. So too does the MAE-values rank-order the LGD-models exactly the same. Lower values of either the KS-statistic or the MAE-value are considered to be better.
Overall, these results suggest that the advanced Type B model is best able to recover the shape and characteristics of the empirical LGD-distribution. We attribute this feat to the innate dichotomisation step that we introduced when rendering the final LGD-estimate per account.

\singlespacing
\printbibliography 
\onehalfspacing



\end{document}